\documentclass[aps,superscriptaddress,twocolumn,pre,longbibliography,floatfix]{revtex4-1}
\usepackage{amsmath,amsthm,amsfonts,amssymb,bm,graphicx,color,mathpazo,times, braket}
\usepackage[colorlinks={true}, citecolor={blue}, filecolor={blue}, linkcolor={blue}, urlcolor={blue}]{hyperref}
\usepackage[caption=false]{subfig}
\usepackage{graphicx}
\usepackage{graphicx}
\usepackage{soul}
\usepackage{appendix}
\usepackage{url}
\usepackage{amsmath}

\allowdisplaybreaks

\begin{document}
	\title{Collective-dissipation-induced dark and metastable-like states for enhanced quantum battery performance}
	\author{Achraf Khoudiri}
	\email{khoudiri.achraf@etu.uae.ac.ma}
	\affiliation{Laboratory of R\&D in Engineering Sciences, Faculty of Sciences and Techniques Al-Hoceima, Abdelmalek Essaadi University, Tetouan, Morocco.}
	\affiliation{
		The UM6P Vanguard Center, Mohammed VI Polytechnic University (UM6P),\\
		Rocade Rabat-Salé, Technopolis,11103, Morocco.}
	\author{Asghar Ullah}
	\email{aullah21@ku.edu.tr}
	\affiliation{Department of Physics, Ko\c{c} University, 34450 Sar\i yer, Istanbul, T\"urkiye}
	
	\author{Abderrahim El Allati}
	\affiliation{Laboratory of R\&D in Engineering Sciences, Faculty of Sciences and Techniques Al-Hoceima, Abdelmalek Essaadi University, Tetouan, Morocco.}
	
	\author{\"Ozg\"ur E. M\"ustecapl\i o\u glu}	
	%\email{omustecap@ku.edu.tr}
	\affiliation{Department of Physics, Ko\c{c} University, 34450 Sar\i yer, Istanbul, T\"urkiye}
	\affiliation{T\"UBITAK Research Institute for Fundamental Sciences (TBEA), 41470 Gebze, T\"urkiye}
	%---------------------------------------------------%
	\date{\today}
	
	\begin{abstract}
		We investigate the role of symmetry-protected dark states and metastable-like frozen states in the autonomous charging dynamics of open quantum batteries described by a transverse-field Ising model. By comparing local and collective dissipation over a range of system sizes, temperatures, and magnetic phases, we demonstrate that collective dissipation generates symmetry-protected dark states together with a much larger set of frozen (metastable) states, forming an extended protected Hilbert space. 
		We derive the multiplicity of the collective dark sector analytically, showing that it follows the Catalan sequence for even system sizes, while such states are absent for odd sizes. Our results show that collective dissipation can enhance ergotropy and charging power, with its advantage depending on temperature, magnetic phase, and system size. While the number of dark and frozen states is identical in the ferromagnetic and antiferromagnetic phases, the achievable ergotropy differs substantially because of the different spectral locations of these protected states. 
		In particular, the antiferromagnetic configuration exhibits considerably larger extractable work owing to the favorable positioning of the protected subspaces within the many-body energy spectrum. Finally, we analyze the active Hilbert-space fraction and show that metastable protection provides an effective mechanism for suppressing dissipative losses while preserving efficient charging pathways. These results establish the dark-state and frozen-state sectors as key resources for optimizing the performance of open quantum batteries through engineered dissipation.
		
	\end{abstract}
	
	%These subspaces host dark states that fundamentally reconfigure energy transfer pathways by shielding the system from environmental degradation.
	\maketitle
	\section{Introduction}
	Recent advances in quantum technologies have enabled precise control of physical systems at the nanoscale, where quantum phenomena strongly influence device behavior~\cite{binder2018thermodynamics, Kosloff, PhysRevE.87.042123, PhysRevLett.105.130401,Alicki1979}. These developments have motivated the design of quantum devices that can store, process, and transfer energy in ways unattainable by classical systems~\cite{Alicki1979, PhysRevB.98.205423, PhysRevB.99.205437, Binder2018, Chenu2019workstatistics}. Among them, quantum batteries have emerged as promising prototypes of energy storage devices that exploit uniquely quantum resources.
	The concept of a quantum battery—a quantum-mechanical device capable of storing and delivering energy—has attracted significant attention in recent years as a potential component of future quantum technologies~\cite{QUACH20232195}. In contrast to classical batteries, quantum batteries can harness uniquely quantum resources, such as coherence, entanglement, and non-classical correlations, to enhance charging speed and energy storage capacity~\cite{PhysRevE.87.042123, RevModPhys.96.031001,PhysRevLett.120.117702, Campaioli2018, Binder_2015, PhysRevLett.129.130602, PhysRevE.102.052109, PhysRevE.97.062105, PhysRevA.106.032212, Hadipure2024,jad2026tight,jad2026suppressing}.

	Most theoretical studies of quantum batteries have so far focused on closed systems, where the dynamics are purely unitary and governed solely by the internal Hamiltonian of the battery together with an external charging field~\cite{PhysRevLett.118.150601, PhysRevB.98.205423, Binder2018}. This idealized framework provides important benchmarks for defining the fundamental limits of stored energy, ergotropy, and charging power~\cite{Kamin2020, PhysRevE.104.054117}. However, in realistic implementations, quantum batteries inevitably interact with their surrounding environment. Such interactions render the dynamics open, leading to dissipation, decoherence, and thermal noise~\cite{Harcoche, weiss2012quantum, Breuer}. Consequently, considerable attention has recently been devoted to open quantum batteries~\cite{PhysRevE.100.032107, PhysRevE.101.062114, PhysRevB.100.115142, PhysRevA.100.043833, PhysRevLett.122.210601, PhysRevResearch.2.013095, PhysRevB.99.035421, PhysRevLett.125.040601, Kamin2020}, whose dynamics may exhibit either Markovian or non-Markovian behavior~\cite{RevModPhys.88.021002, PhysRevLett.103.210401, PhysRevA.86.044101, PhysRevA.90.052118}. In this context, it has been shown that work extraction and charging performance can be significantly enhanced in the presence of collective baths~\cite{mayo2022collective,li2025collective,zhang2024quantum}.
	
	The decoherence-free subspace (DFS) is an important phenomenon in open quantum systems, corresponding to a subspace whose states are protected against environmentally induced decoherence effects \cite{PhysRevLett.81.2594,lidar2003decoherence}. The existence of DFSs has been demonstrated experimentally in both collective and non-collective decoherence processes involving entangled two-qubit polarization states \cite{wu2002creating,kwiat2000experimental}. Such decoherence-protected subspaces are especially relevant for quantum thermodynamic devices, where decoherence directly limits energy-storage performance.
	
	Recently, the role of dark states and DFSs in open quantum batteries has attracted increasing attention. In dissipative many-body systems, dark states may form special components of a broader DFS. In particular, dark states have been exploited to stabilize charging protocols and protect stored energy in dissipative quantum batteries \cite{PhysRevApplied.14.024092}. A central challenge in open quantum batteries is the preservation of extractable work against irreversible environmental losses. Moreover, dissipative environments themselves can,  under suitable conditions, assist charging through dissipative quench protocols \cite{grazi2026charging}. Although previous studies have shown that collective dissipation can enhance charging performance and ergotropy, the microscopic role of interaction-induced dark states in protecting extractable work remains insufficiently understood.
	
	In particular, previous studies have primarily focused on exact dark states, while the broader class of metastable protected states that remain dynamically frozen under dissipation has received much less attention. Furthermore, it is still unclear whether the charging enhancement observed under collective dissipation is determined by the number of protected states, their spectral distribution, or the fraction of the Hilbert space that actively participates in the charging process. In this work, we address these questions by investigating autonomous charging in a transverse-field Ising quantum battery composed of $N$ interacting qubits coupled to thermal reservoirs through either local or collective dissipation. The sign of the Ising interaction gives rise to ferromagnetic (FM) and antiferromagnetic (AFM) magnetic phases, allowing us to explore how magnetic ordering influences dissipative charging.

	Our analysis reveals that collective dissipation supports a symmetry-protected family of dark states, together with a substantially larger set of dynamically frozen states that form a metastable-like protected sector of the Hilbert space. We analytically derive the multiplicity of the collective dark sector and show that, for even system sizes, it follows the Catalan sequence, whereas such states are absent for odd $N$. We further examine the invariance of this dark sector under the battery Hamiltonian and show that, in the present model, it constitutes an exact decoherence-free subspace for $N=2$, whereas for larger even system sizes, Hamiltonian-induced leakage prevents exact DFS invariance.\\
	Beyond these symmetry-protected states, we numerically identify dynamically frozen states through the approximate conservation of their energy-level populations, reflecting a dynamical balance between dissipative gain and loss. By comparing the FM and AFM configurations, we show that although they have the same numbers of dark and frozen states for a given system size, their charging performance differs substantially because the protected states occupy different regions of the many-body energy spectrum. Finally, by introducing the active Hilbert-space fraction, we demonstrate that the charging performance is governed by the interplay between dissipative protection and the availability of active energy-transfer pathways, thereby providing a unified framework for understanding and optimizing autonomous quantum batteries.
	
	The remainder of this paper is organized as follows. In Sec.~\ref{model}, we introduce the transverse-field Ising quantum battery, the local and collective dissipative models, and the performance measures. Section~\ref{results} presents the characterization of dark states, frozen states, and the corresponding protected Hilbert-space structure, followed by an analysis of their impact on ergotropy and charging power. The dependence on the system size is discussed in Sec.~\ref{Sec:Scaling}. Finally, Sec.~\ref{conc} summarizes our main findings.
	%***************************************************%
	\section{Model and Dissipative Dynamics}\label{model}
	\subsection{Quantum battery model}
	%***************************************************%
	We consider a quantum battery described by the transverse-field Ising model in the presence of an external magnetic field. The quantum battery Hamiltonian is given by
	\begin{equation}
		H_B = - J \sum_{i=1}^{N-1} \sigma_i^z \sigma_{i+1}^z - h \sum_{i=1}^{N} \sigma_i^x,
	\end{equation}
	where \(J\) is the spin-spin interaction strength, \(h\) denotes the strength transverse field, and \(\sigma_i^{x,z}\) are the Pauli operators acting on site \(i\).

	For a general system of $N$ qubits, the battery Hamiltonian $H_B$ acts on a Hilbert space of dimension $2^N$. The Hamiltonian can be diagonalized in this finite-dimensional space. Denoting its eigenstates and eigenvalues by $\{|E_k^{\alpha}\rangle\}$ and $\{E_k^{\alpha}\}$, respectively, where $\alpha \in \{\mathrm{AFM}, \mathrm{FM}\}$ corresponds to the antiferromagnetic (AFM) case $(J<0)$ and the ferromagnetic (FM) case $(J>0)$, we write
	\begin{equation}\label{Diagonalized_Hamiltonian}
		H_B^{\alpha} = \sum_{k=0}^{2^N -1} E_k^{\alpha} \, |E_k^{\alpha}\rangle \langle E_k^{\alpha}|.
	\end{equation}
	
	The corresponding eigenstates $\{|E_k^{\alpha}\rangle\}$ and eigenvalues $\{E_k^{\alpha}\}$ are obtained by diagonalizing $H_B^{\alpha}$ in the computational basis
	\begin{equation}
		\left\{ |s_1 s_2 \cdots s_N\rangle \right\},
		\qquad s_i \in \{\uparrow,\downarrow\}.
	\end{equation}
	
	In what follows, we work in the eigenbasis $\{|E_k^{\alpha}\rangle\}$ of $H_B^{\alpha}$. The transformation from the computational basis $\{|s_1 s_2 \cdots s_N\rangle\}$ to the eigenbasis $\{|E_k^{\alpha}\rangle\}$ is described by a unitary matrix $U^{\alpha}$. Accordingly, any operator $O$ can be expressed in the eigenbasis as
	\begin{equation}
		O' = \left(U^{\alpha}\right)^{\dagger} O U^{\alpha}.
	\end{equation}

	%***********************************************
	\subsection{Open-system dynamics: Local and collective dissipation}\label{sec:open_system}
	%***********************************************
	We consider the quantum battery as an open quantum system weakly coupled to a thermal bath. The dynamics of the system density matrix $\rho$ is described within the Markovian approximation by the Gorini--Kossakowski--Sudarshan--Lindblad (GKSL) master equation~\cite{Breuer},
	\begin{equation}
		\frac{d\rho}{dt}= -i[H,\rho] + \sum_{\mu}
		\gamma_\mu \, \mathcal{D}[L_\mu]\rho,
	\end{equation}
	where $H = H_B$ is the battery Hamiltonian, $L_\mu$ are the jump operators describing system--bath interactions, and $\mathcal{D}[O]\rho$ is the Lindblad dissipator defined as
	\begin{equation}
		\mathcal{D}[O]\rho = O\rho O^\dagger - \frac{1}{2} \left\{ O^\dagger O,\rho \right\}.
	\end{equation}
	The thermal excitation and relaxation processes are characterized by the rates
	\begin{equation}
		\gamma_{\downarrow} = \gamma (n_{\mathrm{th}} + 1), 
		\qquad
		\gamma_{\uparrow} = \gamma n_{\mathrm{th}},
	\end{equation}
	where $n_{\mathrm{th}} = (e^{\omega_0/T} - 1)^{-1}$ is the mean thermal occupation number at temperature $T$, and $\omega_0$ denotes the characteristic transition frequency of the system. Here, $T$ is interpreted as an effective temperature parameter of an engineered incoherent pump-and-decay reservoir, rather than the thermodynamic temperature of a bath in equilibrium with the interacting battery Hamiltonian $H_B$. Accordingly, the resulting long-time state is referred to as a stationary state rather than a thermal-equilibrium state. We further note that, since the Markovian rates are fixed and independent of the number of qubits, the reservoir is not depleted, and finite-resource effects are not captured in our model.
	\paragraph*{Local dissipation---}
	In the case of local system--bath coupling, each qubit interacts independently with its own environment. The corresponding jump operators are given by the local spin lowering and raising operators,
	\begin{equation}\label{OPINT_LOC}
		L_i = \sigma_i^-,
		\qquad
		L_i^\dagger = \sigma_i^+,
	\end{equation}
	leading to the dissipative contribution
	\begin{equation}
		\mathcal{L}_{\mathrm{loc}}[\rho] = \sum_{i=1}^N
		\left[ \gamma_{\downarrow} \, \mathcal{D}[L_i]\rho + \gamma_{\uparrow} \, \mathcal{D}[L_i^\dagger]\rho
		\right].
	\end{equation}
	This captures incoherent energy exchange with independent baths and typically leads to relaxation toward a product thermal state. 
	\paragraph*{Collective dissipation---} In contrast, for collective coupling all qubits interact with a common bath, giving rise to correlated decay and excitation processes. The jump operators are defined in terms of collective spin operators,
	\begin{equation}\label{OPINT_COL}
		L = \sum_{i=1}^N \sigma_i^-,
		\qquad
		L^\dagger = \sum_{i=1}^N \sigma_i^+,
	\end{equation}
	and the corresponding dissipator takes the form
	\begin{equation}
		\mathcal{L}_{\mathrm{col}}[\rho] = \gamma_{\downarrow} \, \mathcal{D}[L]\rho +
		\gamma_{\uparrow} \, \mathcal{D}[L^\dagger]\rho.
	\end{equation}
	Such collective interactions induce correlated quantum jumps and can lead to phenomena such as superradiance and enhanced coherence. Hence, the system evolution is governed either by local or collective dissipation, leading to the full master equation
	\begin{equation}
		\frac{d\rho}{dt} = -i[H,\rho]
		+ \mathcal{L}_{\mathrm{loc}}[\rho]
		\quad \text{or} \quad
		-i[H,\rho] + \mathcal{L}_{\mathrm{col}}[\rho].
	\end{equation}
	This framework allows us to systematically investigate how different environmental couplings affect the charging dynamics and performance of the quantum battery. In our case, the energy is injected into the system through the engineered dissipative channels rather than via an external driving field.
	
	%********************************************************************%
	\subsection{Performance metrics}
	%********************************************************************%
	
	The ergotropy quantifies the maximum amount of work that can be extracted from a quantum state $\rho$ through unitary operations with respect to the battery Hamiltonian $H_B$~\cite{Allahverdyan_2004}. Let $\{E_k^{\alpha}\}$ denote the eigenvalues of $H_B^{\alpha}$, arranged in nondecreasing order, i.e., $E_{k+1}^{\alpha} \geq E_k^{\alpha}$ for $k=0,\ldots,2^N-1$, and let $\{P_k\}$ be the eigenvalues of the density matrix $\rho$, sorted in nonincreasing order, i.e., $P_{k+1} \leq P_k$. The passive energy associated with the state $\rho$ is defined as
	\begin{equation}\label{Eq.Stored_Energy_passive}
		E_{\mathrm{passive}}(\rho) = \sum_{k=0}^{2^N-1} P_k E_k^{\alpha}.
	\end{equation}
	
	The energy stored in the battery is given by
	\begin{equation}\label{Eq.Stored_Energy}
		E(\rho) = \mathrm{Tr}(\rho H_B^{\alpha}).
	\end{equation}
	
	The ergotropy $\mathcal{W}$ is defined as the difference between the stored energy and the passive energy,
	\begin{equation}\label{Eq.ERGOTROPY}
		\mathcal{W}(\rho) = E(\rho) - E_{\mathrm{passive}}(\rho).
	\end{equation}
	
	The charging power, denoted by $\mathcal{P}$, describes the rate at which the battery stores useful extractable work during the charging process. Mathematically, it is defined as
	\begin{equation}
		\mathcal{P} = \frac{\mathcal{W}(\rho)}{t}.
	\end{equation}
	
	%In our model, the ergotropy is determined by the difference between the stored energy $E(\rho)$ and the passive energy $E_{\mathrm{passive}}(\rho)$. 
	The passive state is constructed by rearranging the populations in decreasing order according to the increasing order of the Hamiltonian eigenvalues. In this context, the presence of dark states in both the AFM and FM configurations under collective dissipation, and their absence under local dissipation, plays a central role. Since dark states are unaffected by reservoir-induced decoherence, we analyze how their presence under collective dissipation, compared to their absence under local dissipation, impacts the ergotropy performance of the quantum battery.
	
	%********************************************************************%
	\section{Dark States and Decoherence-Free Subspaces}
	\label{results}
	%********************************************************************%
	
	A decoherence-free subspace (DFS) is a sector of the system Hilbert space in which quantum states are immune to a given environmental noise. Such subspaces naturally arise when the system--bath interaction exhibits specific symmetries.
	
	Let $\mathcal{H}$ denote the total Hilbert space. A DFS is a subspace
	$\widetilde{\mathcal{H}} \subseteq \mathcal{H}$ spanned by a set of basis states
	$\{\ket{\widetilde{E}_k}\} \subset \widetilde{\mathcal{H}}$, such that any state of the form
	\begin{equation}
		\widetilde{\rho}
		=
		\sum_{k,k'}
		\widetilde{\rho}_{kk'}
		\ket{\widetilde{E}_k}
		\bra{\widetilde{E}_{k'}}
	\end{equation}
	remains unaffected by decoherence, namely,
	\begin{equation}
		\mathcal{L}[\widetilde{\rho}] = 0.
	\end{equation}
	
	This condition is satisfied when the Lindblad operators $L_{\mu}$ act proportionally on all basis states within the subspace,
	\begin{equation}
		\label{DFS_CONDITION}
		L_{\mu}\ket{\widetilde{E}_k}
		=
		c_{\mu}\ket{\widetilde{E}_k},
		\qquad
		\forall\, k,\mu,
	\end{equation}
	where $c_{\mu}$ is independent of $k$. In the particular case where
	$c_{\mu}=0$, the states are annihilated by the system--bath coupling operators and are referred to as \emph{dark states}.
	
	We introduce the projector onto the dark subspace,
	\begin{equation}
		\widetilde{\Pi}
		=
		\sum_k
		\ket{\widetilde{E}_k}
		\bra{\widetilde{E}_k}.
	\end{equation}
	Although the collective dissipator vanishes on this subspace, its invariance under the full quantum-battery Hamiltonian requires
	\begin{equation}\label{Condition_of_invariant}
		\left(
		\mathbb{I}
		-
		\widetilde{\Pi}
		\right)
		H
		\widetilde{\Pi}
		=
		0,
	\end{equation}
	where $\mathbb{I}$ denotes the identity operator on the total Hilbert space. Under these conditions, the environment cannot distinguish between different states within the subspace, and quantum coherence is therefore preserved throughout the evolution. Consequently, any state initialized within the DFS remains protected from decoherence.

	%********************************************************************%
	\subsection{Exact Dark-State Conditions}
	%********************************************************************%
	
	For local dissipation, each qubit $i$ is directly coupled to its own reservoir through the interaction operators given in Eq.~\eqref{OPINT_LOC}. Physically, this means that the interaction induces transitions in the computational basis
	\begin{equation}
		\left\{
		\ket{s_1 s_2 \cdots s_N}
		\right\}
	\end{equation}
	of the qubits, where the action of the local operators generally maps the states outside the DFS, such that
	\begin{equation}
		L_i \ket{E_k^\alpha}
		=
		\ket{\psi},
		\qquad
		\ket{\psi}
		\notin
		\widetilde{\mathcal{H}}.
	\end{equation}
	
	For collective dissipation, the set of $N$ qubits is collectively coupled to the reservoir through the interaction operators given in Eq.~\eqref{OPINT_COL}. In this case, the total Hilbert space $\mathcal{H}_B$ of the $N$ qubits can be decomposed as a direct sum over the total-spin sectors, denoted by $S$, according to the decomposition given in Eq.~\eqref{DEC_Hilbert}. Moreover, any quantum state $\ket{\psi}$ can be expressed in the total-spin basis, as shown in Eq.~\eqref{anypsi}. The DFS condition in Eq.~\eqref{DFS_CONDITION}, together with Eq.~\eqref{SM}, is satisfied for strict dark states obeying
	\begin{align}
		L\ket{\psi} &= 0,\\
		L^\dagger\ket{\psi} &= 0.
	\end{align}
	
	Therefore, the existence of a strict dark state requires
	$S_{\min}=0$, which is possible only when the number of qubits is even, i.e.,
	\begin{equation} \label{N_even}
		N = 2n,
		\qquad
		n \in \mathbb{N}^{*}.
	\end{equation}
	For odd $N$, one has $S_{\min}=1/2$, which prevents the formation of strict dark states under these collective symmetric operations.
	
	Physically, the presence of strict dark states originates from the $SU(2)$ symmetry associated with the total spin of all qubits in the quantum battery. This corresponds to the DFS condition in Eq.~\eqref{DFS_CONDITION}, which follows from the symmetric behavior imposed by the collective interaction operators given in Eq.~\eqref{OPINT_COL}. The condition $S=M=0$ implies that the local spins of the qubits cancel each other, resulting in a vanishing total spin of the quantum battery. In this case, the battery remains in a collective equilibrium of the qubit spins. However, the existence of strict dark states is a consequence of symmetry and is not a thermal phenomenon; rather, it is an internal property of the quantum-battery model and its collective-spin structure.

	Although the collective dissipator vanishes on the dark-state subspace, this does not by itself guarantee invariance under the full Hamiltonian. Following the criterion introduced in Eq.~\eqref{Condition_of_invariant}, we therefore quantify the Hamiltonian-induced leakage from this subspace through
	\begin{equation}
		\eta
		=
		\frac{
			\left\|
			\left(
			\mathbb{I}
			-
			\widetilde{\Pi}
			\right)
			H_B
			\widetilde{\Pi}
			\right\|_F
		}{
			\left\|
			H_B
			\widetilde{\Pi}
			\right\|_F
		}
		<
		\varepsilon,
		\qquad
		\varepsilon \rightarrow 0.
	\end{equation}
	Throughout this work, the numerical tolerance is chosen as
	$\varepsilon=10^{-9}$. Here, $\|\cdot\|_F$ denotes the Frobenius norm, defined for an arbitrary operator $A$ as
	\begin{equation}
		\|A\|_F
		=
		\sqrt{
			\operatorname{Tr}
			\left(
			A^\dagger A
			\right)
		}.
	\end{equation}
	The quantity $\eta$ quantifies the relative leakage of the dark-state subspace and its invariance under the quantum-battery Hamiltonian. Therefore, dark states satisfying the collective jump-operator conditions in Eq.~\eqref{OPINT_COL} are exact DFS states when the corresponding Hamiltonian leakage satisfies $\eta<\varepsilon$; otherwise, they remain dark states with respect to the collective dissipative interaction operators.
	
	\begin{figure}[t!]
		\centering
		\includegraphics[width=1\linewidth]{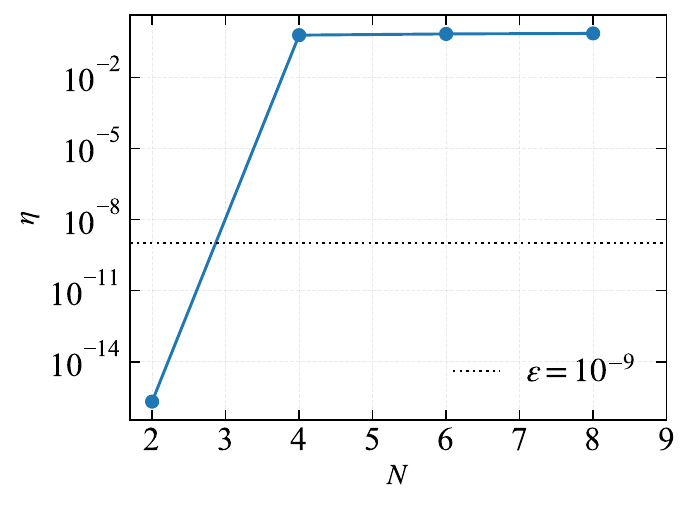}
		\caption{Normalized Hamiltonian leakage parameter $\eta$ as a function of the number of qubits $N$. The dotted line indicates the invariance tolerance $\varepsilon=10^{-9}$. Similar behavior is obtained for both the FM ($J=1$) and AFM ($J=-1$) cases.}
		\label{fig:DFS_condition_eta}
	\end{figure}
	
	Figure~\ref{fig:DFS_condition_eta} shows the normalized Hamiltonian leakage parameter $\eta$ of the quantum battery Hamiltonian $H_B  $ as a function of the number of qubits $N$. Similar behavior is observed for both the FM and AFM cases under collective dissipation. For $N=2$, the value of $\eta$ lies well below the numerical tolerance $\varepsilon=10^{-9}$, indicating that the dark subspace is invariant under the battery Hamiltonian and therefore constitutes an exact decoherence-free subspace. In contrast, for $N\geq4$, $\eta$ exceeds the invariance threshold, indicating that the dark subspace is not invariant under $H_B$
	and therefore does not constitute an exact DFS of the full dynamics. Thus, while the $N=2$ dark state forms an exact DFS, for larger system sizes the identified states remain dark with respect to the collective jump operators but are subject to Hamiltonian-induced leakage from the dark subspace. Hereafter, we use the term “dark states” to refer to states that are dark with respect to the collective jump operators, unless otherwise specified.

	%********************************************************************%
	\subsection{Number of  dark states}
\begin{figure*}[t!]
		\centering
		\includegraphics[scale=0.95]{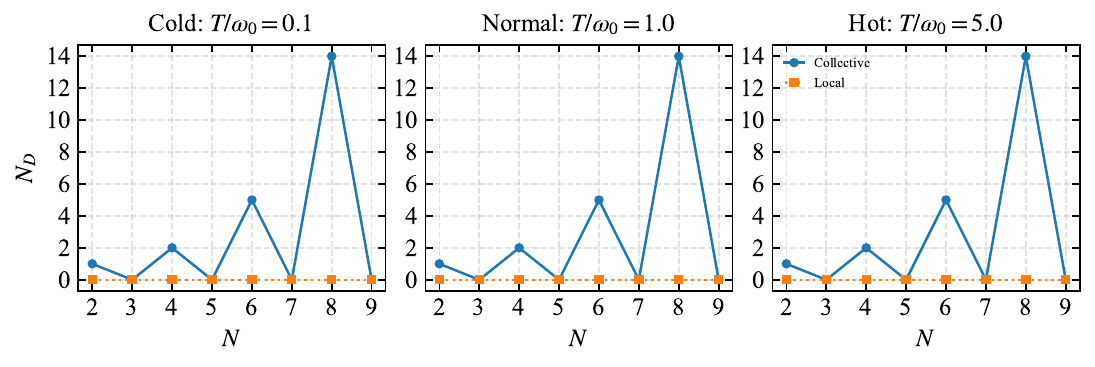}
		\includegraphics[scale=0.95]{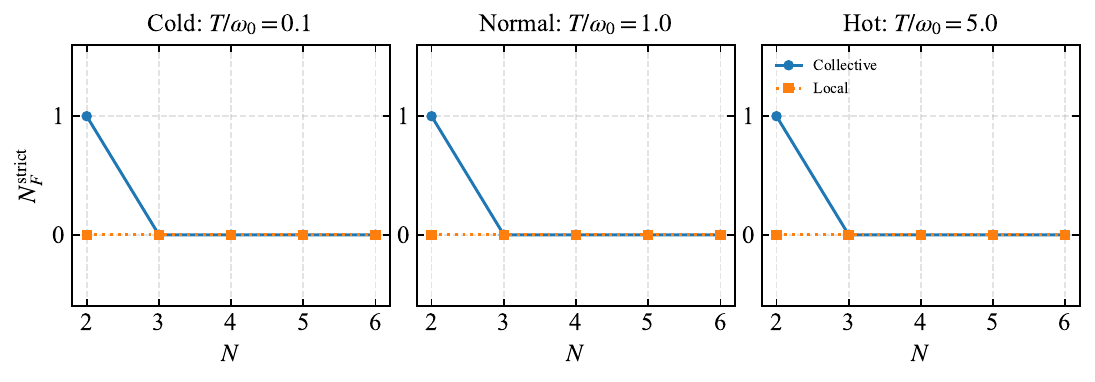}
		\includegraphics[scale=0.95]{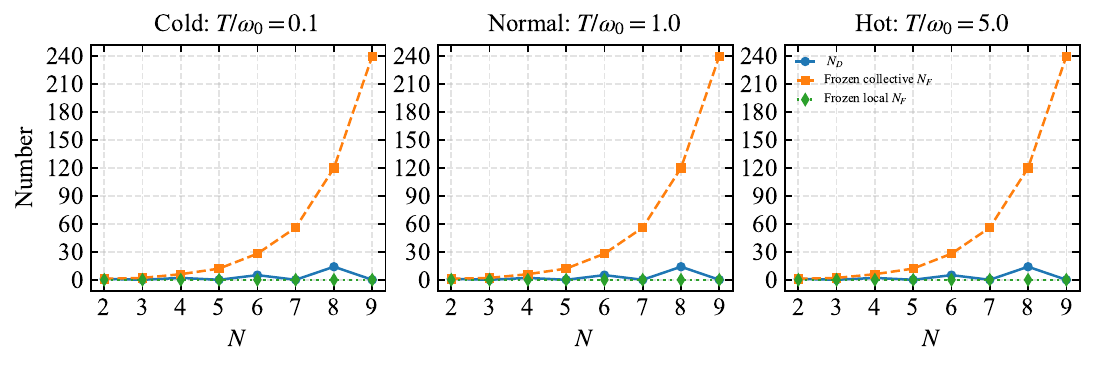}
		\caption{
			\textbf{Top panel:} Number of dark states $N_D$ as a function of the number of qubits $N$ for three bath-temperature regimes: cold bath ($T/\omega_0=0.1$), normal bath ($T/\omega_0=1$), and hot bath ($T/\omega_0=5$), respectively, for collective and local dissipation. We set $J=1$ and $B=0.1$.
			\textbf{Middle panel:} Number of strict frozen states $N_F^{\mathrm{strict}}$ as a function of the number of qubits $N$ for the same three bath-temperature regimes, for collective and local dissipation. We set $J=1$ and $h=0.1$.
			\textbf{Bottom panel:} Number of frozen states $N_F$ and dark states $N_D$ as a function of the number of qubits $N$ for the same three bath-temperature regimes, for collective and local dissipation. We set $J=1$ and $h=0.1$.
		}
		\label{fig:Dark_Frozen_States}
	\end{figure*}

	%********************************************************************%
	%jump-dark states: let us use dark states for our case, coming jump operator
	Let the number of dark states be denoted by $N_D$. As discussed in Appendix~\ref{Der_Dark_state}, the number of  dark states is given by the multiplicity of the total-spin sector $S=0$,
	\begin{align}
		N_D = d_0 .
	\end{align}
	For $N$ spin-$1/2$ particles, the multiplicity of the total-spin sector $S$ is denoted by $d_S$ and is given by Eq.~\eqref{ds_multi}. Since dark states are present only when $N$ is even, according to the condition in Eq.~\eqref{N_even}, the number of dark states is given by Eq.~\eqref{DSNAP}. These states are quantified by the Catalan number, denoted by $C_n$, and is given as
	\begin{equation}
		C_n=\frac{1}{n+1}\binom{2n}{n}, 
	\end{equation}
	as demonstrated in Eq.~\eqref{CATALAN}. Since $n=N/2$, the number of strict dark states is therefore given by
	\begin{equation}
		N_D = C_{\frac{N}{2}}.
	\end{equation}		
	Numerically, we identify the dark states in our scenario using using the condition in Eq.~\eqref{DFS_CONDITION}, for three reservoir conditions: a cold bath ($T/\omega_0 = 0.1$), a normal bath ($T/\omega_0 = 1$), and a hot bath ($T/\omega_0 = 5$), respectively, for collective and local dissipation, as shown in Fig.~\ref{fig:Dark_Frozen_States} (top panel). As the number of qubits increases, the number of dark states increases for even $N$. In particular, it increases according to the Catalan numbers $(1,2,5,14)$ for even $N=(2,4,6,8)$, while dark states are absent for odd $N=(3,5,7)$. The numerical results agree with the theoretical results. Note that, even under temperature variations, the number of dark states remains the same because the dark states originate directly from spin symmetry.
	
	In the case of local dissipation, each qubit interacts independently with its own environment, and the dynamics is governed by a set of local Lindblad operators $\{L_i\}$ acting on each qubit $i$. In contrast, for collective dissipation, the coupling is described by a single global operator of the form $L = \sum_{i=1}^{N} L_i$, reflecting the identical interaction of all qubits with a common bath.
	
	The lack of symmetry in the local-dissipation case prevents the environment from acting uniformly on the system, thereby destroying the conditions required for the presence of dark states and DFS-like sectors. Instead, local dissipation induces transitions between the local basis states of each qubit, such as spin flips from $\ket{\downarrow}$ to $\ket{\uparrow}$ and vice versa. These processes lead to the loss of quantum coherence and inhibit the formation of a decoherence-free-like subspace.
	
	%********************************************************************%
	\section{Metastable-like frozen states and the protected sector}
	%********************************************************************%
	%----------------------------------------------------------------%
	%----------------------------------------------------------------%		
	It is important to emphasize that a \emph{frozen state} is not necessarily equivalent to a dark state. We define a frozen energy level $\ket{E_k}$ as a level whose population remains approximately constant during the system evolution.
	
	Let us first introduce the projector associated with the energy eigenstate $\ket{E_k}$,
	\begin{equation}
		\Pi_k = \ket{E_k}\bra{E_k}.
	\end{equation}
	A \emph{strict frozen state} is defined by the condition
	\begin{equation}\label{Strict_frozen_states_condition}
		\mathcal{L}[\Pi_k] = 0.
	\end{equation}
	This definition is independent of the initial state and depends only on the physical properties of the system and its dissipative dynamics. More generally, frozen states can be defined dynamically through the approximate conservation of the corresponding level population
	\begin{equation}
		P_k(t) = \langle E_k | \hat{\rho}(t) | E_k \rangle \simeq \mathrm{const.},
	\end{equation}
	or, equivalently,
	\begin{equation}\label{F_states}
		\frac{d}{dt} P_k(t) = \frac{d}{dt} \langle E_k | \hat{\rho}(t) | E_k \rangle
		= \langle E_k | \dot{\hat{\rho}}(t) | E_k \rangle \simeq 0.
	\end{equation}
	The system dynamics are governed by the master equation
	\begin{equation}
		\frac{d}{dt}\hat{\rho}(t) = -i[\hat{H}_B,\hat{\rho}(t)] + \mathcal{L}[\hat{\rho}(t)],
	\end{equation}
	where the dissipative contribution is given by
	\begin{equation}
		\mathcal{L}[\hat{\rho}(t)] = \sum_{\mu} \gamma_{\mu} \mathcal{D}[L_{\mu}]\big[\hat{\rho}(t)\big].
	\end{equation}
	Substituting the dissipator $\mathcal{D}[L_{\mu}]\big[\hat{\rho}\big]$ into the population dynamics gives
	\begin{equation}
		\gamma_{\mu} \mathcal{D}[L_{\mu}][\hat{\rho}]
		= \underbrace{\gamma_{\mu} L_{\mu}\hat{\rho}L_{\mu}^{\dagger}}_{\text{population gain}}
		- \underbrace{\frac{\gamma_{\mu}}{2} \left\{ L_{\mu}^{\dagger}L_{\mu}, \hat{\rho} \right\}}_{\text{population loss}}.
	\end{equation}
	%-------------------------------------------------------------------%
	\begin{figure*}[t!]
		\centering
		\includegraphics[scale=0.97]{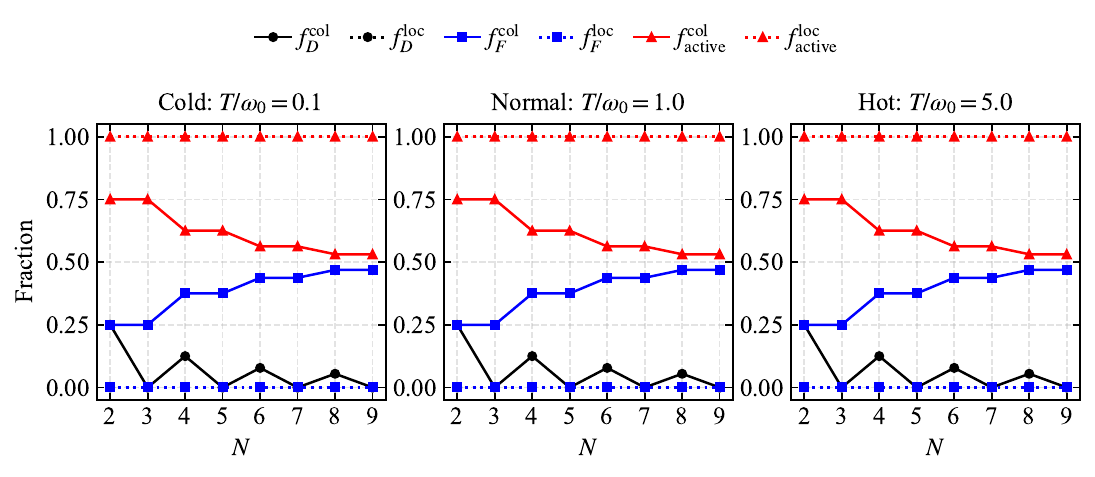}
		\caption{Fractions $f_{D(F)}^{\mathrm{loc}(\mathrm{col})}$ of frozen and dark states and the active-state fraction $f_{\mathrm{active}}^{\mathrm{loc}(\mathrm{col})}$ as functions of the number of qubits $N$ for three bath-temperature scenarios: cold bath ($T/\omega_0 = 0.1$), normal bath ($T/\omega_0 = 1$), and hot bath ($T/\omega_0 = 5$), respectively, for collective and local dissipation. We set $J=1$ and $h=0.1$.}
		\label{Fraction_Frozn_DARK_STATES_figure}
	\end{figure*}
	%-------------------------------------------------------------------%
	Therefore, the condition for a frozen state can be analyzed by requiring that the gain and loss contributions balance each other in the population equation for the level $\ket{E_k}$,
	\begin{equation}
		\langle E_k | \dot{\hat{\rho}}(t) | E_k \rangle \simeq 0.
	\end{equation}
	Physically, this condition is satisfied when
	\begin{align}
		W_{\mathrm{gain},k} &= W_{\mathrm{loss},k},\\
		W_{\mathrm{gain},k} &= \sum_{\mu} \langle E_k | \gamma_{\mu} L_{\mu}\hat{\rho}L_{\mu}^{\dagger} | E_k \rangle,\\
		W_{\mathrm{loss},k} &= \sum_{\mu} \left\langle E_k \left| \frac{\gamma_{\mu}}{2}
		\left\{ L_{\mu}^{\dagger}L_{\mu}, \hat{\rho} \right\} \right| E_k \right\rangle .
	\end{align}
	The frozen states identified here are not strict frozen states. Rather, they form a metastable-like protected subspace within the total Hilbert space, in which the population-gain and population-loss contributions remain approximately balanced. These states are not necessarily dark states of the reservoir; instead, they correspond to energy levels whose populations remain approximately constant because the gain and loss processes remain nearly balanced over the considered evolution time.

	%********************************************************************%
	\subsection{Number of Frozen States and Strict Frozen States}
	%********************************************************************%	
	We numerically quantify the number of frozen states using the condition introduced in Eq.~\eqref{F_states}. The number of frozen states, denoted by $N_F$, is determined according to
	\begin{equation}
		\max_t
		\left|
		\frac{dP_k(t)}{dt}
		\right|
		<
		\varepsilon,
		\qquad
		\varepsilon \rightarrow 0.
	\end{equation}
	For strict frozen states, we employ the condition given in Eq.~\eqref{Strict_frozen_states_condition}. The number of strict frozen states, denoted by $N_F^{\mathrm{strict}}$, is determined numerically according to
	\begin{equation}\label{strict_frozen}
		\left\|
		\mathcal{L}[\Pi_k]
		\right\|
		<
		\varepsilon,
		\qquad
		\varepsilon \rightarrow 0.
	\end{equation}
	
	Throughout this work, the convergence tolerance is chosen as
	$\varepsilon = 10^{-9}$. In Fig.~\ref{fig:Dark_Frozen_States} (middle panel), we compute the number of strict frozen states, independently of the initial state of the quantum battery, using the condition given in Eq.~\eqref{strict_frozen}. We consider the three bath-temperature regimes, namely, cold, normal, and hot baths, and compare collective and local dissipation. For collective dissipation, strict frozen states are observed only for $N=2$, whereas for $N>2$ no strict frozen states are found. In the case of local dissipation, strict frozen states are absent for all considered values of the qubit number $N$. Physically, this indicates the absence of metastable-subspace protection under local dissipation, while for collective dissipation such protection occurs only for $N=2$, owing to the presence of an exact DFS in this case.
	
	We compute the number of frozen states number in Fig.~\ref{fig:Dark_Frozen_States} (bottom panel) and compare it with the number of dark states as a function of the number of qubits $N$, for three reservoir conditions: a cold bath ($T/\omega_0 = 0.1$), a normal bath ($T/\omega_0 = 1$), and a hot bath ($T/\omega_0 = 5$), for both collective and local dissipation. For collective dissipation, the number of frozen states increases exponentially with the number of qubits. Moreover, frozen states are present for both even and odd values of $N$, in contrast to dark states, which appear only for even $N$. The number of frozen states is also significantly larger than the number of dark states. Physically, the presence of frozen states under collective dissipation is associated with the symmetry imposed by the global interaction operators $L$ and $L^\dagger$ in the computational basis and is represented as a metastable-like subspace over time. For local dissipation, the absence of frozen states for all considered values of $N$ is attributed to the individual coupling of each qubit $i$ to its own reservoir, together with the absence of metastable levels.
	
	%--------------------------------------------------------------------%
	\begin{figure*}[t!]
		\centering
		\includegraphics[width=0.98\textwidth]{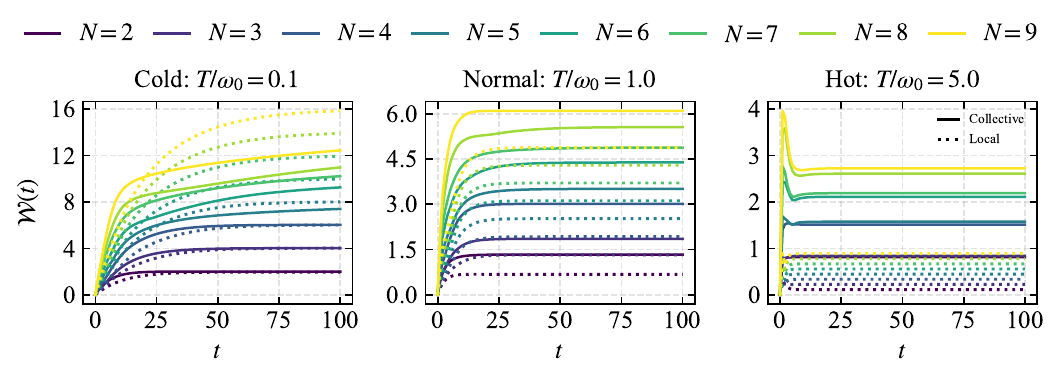}\\
		\includegraphics[width=0.98\textwidth]{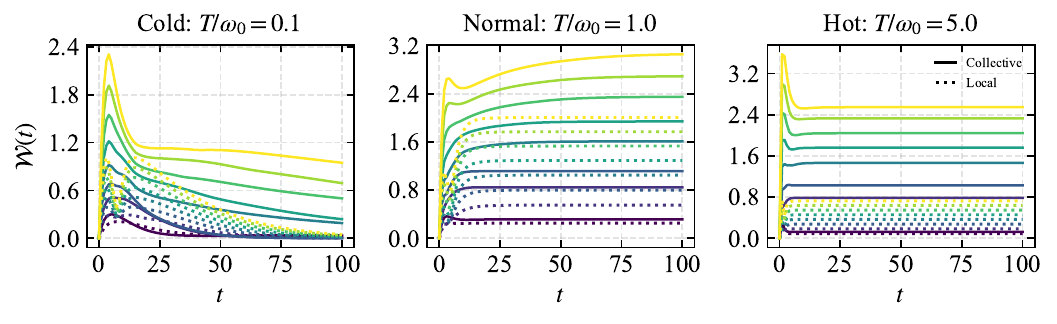}
		\caption{Dynamics of the ergotropy $\mathcal{W}(\rho(t))$ as a function of time $t$ and the number of qubits $N$ for three bath-temperature scenarios: cold bath ($T/\omega_0 = 0.1$), normal bath ($T/\omega_0 = 1$), and hot bath ($T/\omega_0 = 5$), respectively, for collective and local dissipation. The top panel corresponds to AFM coupling ($J=-1$), while the bottom panel corresponds to FM coupling ($J=1$). The solid red curves correspond to collective interactions, while the dotted blue curves correspond to local interactions with the bath. The remaining parameters are set to $h=0.1$, $\gamma=0.09$, and $\omega_0=2h$.}
		\label{ERGOTROPY}
	\end{figure*}
	%--------------------------------------------------------------------%

	%********************************************************************%
	\subsection{Beyond dark and frozen states}
	%********************************************************************%
	
	As discussed above, dark states are protected by the symmetry of the total spin. They are not thermal phenomena but are directly linked to the annihilation of the total spin momentum and to the absence of magnetization, $S=M=0$. They are quantified by Catalan numbers and represent a part of the total Hilbert space denoted by $\tilde{\mathcal{H}}_B \subset \mathcal{H}_B$. Frozen states, in contrast, are dynamically protected by the balance between incoming and outgoing transitions, namely gain and loss, and are described by a part of the Hilbert space denoted by $\bar{\mathcal{H}}_B \subset \mathcal{H}_B$ of the quantum battery $B$.
	
	To better understand how each sector, $\tilde{\mathcal{H}}_B$ and $\bar{\mathcal{H}}_B$, corresponding to dark and frozen states, respectively, contributes to the total Hilbert space $\mathcal{H}_B$, we use the ratio between $N_D$ or $N_F$ and the total number of levels in the Hilbert space, $2^N$. This fraction is denoted by $f_{D(F)}^{\mathrm{loc}(\mathrm{col})}$ and is mathematically defined as
	\begin{equation}
		f_{D(F)}^{\mathrm{loc}(\mathrm{col})} = \frac{N_{D(F)}}{2^N}.
	\end{equation}
	The quantity $f_{D(F)}^{\mathrm{loc}(\mathrm{col})}$ describes the contribution of the dark sector $f_D^{\mathrm{loc}(\mathrm{col})}$ or the frozen sector $f_F^{\mathrm{loc}(\mathrm{col})}$ to the total Hilbert space of the quantum battery for local or collective dissipation.
	%*********************************************************%
	\begin{figure*}[t!]
		\centering
		\includegraphics[width=0.98\textwidth]{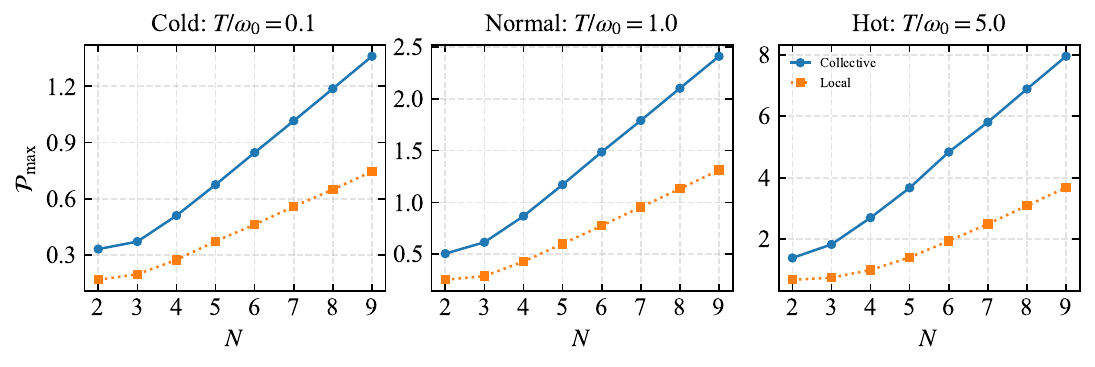}\\
		\includegraphics[width=0.98\textwidth]{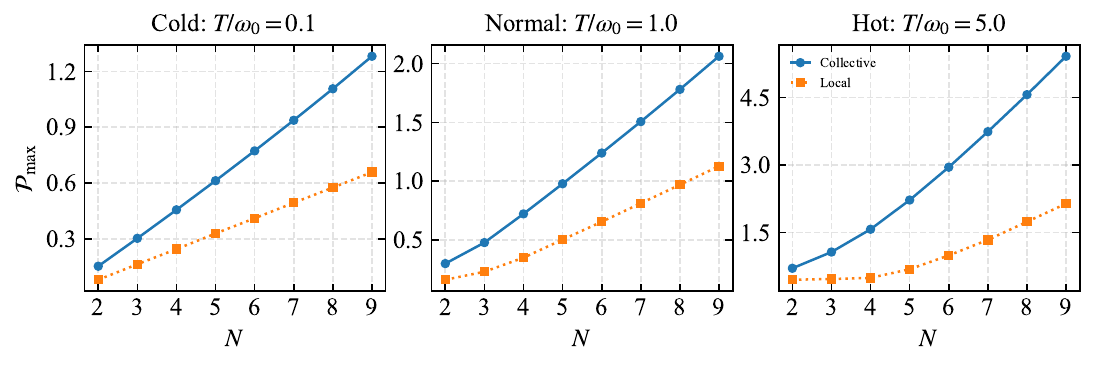}	
		\caption{Maximal charging power $\mathcal{P}_{\max}$ as a function of the number of qubits $N$ for three bath-temperature scenarios: cold bath ($T/\omega_0 = 0.1$), normal bath ($T/\omega_0 = 1$), and hot bath ($T/\omega_0 = 5$), respectively, for collective and local dissipation. The top panel corresponds to AFM coupling ($J=-1$), while the bottom panel corresponds to FM coupling ($J=1$). The solid red curves correspond to collective interactions, while the dotted blue curves correspond to local interactions with the bath. The remaining parameters are set to $h=0.1$, $\gamma=0.09$, and $\omega_0=2h$.}
		\label{PMAX}
	\end{figure*}
	%********************************************************%
	In Fig.~\ref{Fraction_Frozn_DARK_STATES_figure}, we plot the fractions
	$f_{D(F)}^{\mathrm{loc}(\mathrm{col})}$ of dark and frozen states as a function
	of the number of qubits $N$ for three reservoir temperatures: a cold bath
	($T/\omega_0 = 0.1$), an intermediate-temperature bath ($T/\omega_0 = 1$), and
	a hot bath ($T/\omega_0 = 5$), under both collective and local dissipation. For local dissipation, the contributions of both frozen and dark states are zero. For collective dissipation, the fraction of frozen states increases with the number of qubits, in contrast to the dark-state fraction, which decreases with increasing $N$. This means that the contribution of the frozen sector $\bar{\mathcal{H}}_B$ is higher than that of the dark-state sector $\tilde{\mathcal{H}}_B$. It is also clear that the dark-state sector is part of the frozen sector, while the frozen sector is part of the total Hilbert space
	\begin{equation}
		\tilde{\mathcal{H}}_B \subset\mathcal{H}_B, \quad  \bar{\mathcal{H}}_B \subset\mathcal{H}_B  .
	\end{equation}
	
	Note that when the number of frozen states and dark states increases, the number of active levels in the total Hilbert space $\mathcal{H}_B$ decreases. Let the number of active states be denoted by
	\begin{equation}
		N_{\mathrm{active}} = 2^N - N_F .
	\end{equation}
	The fraction of active states, denoted by $f_{\mathrm{active}}^{\mathrm{loc}(\mathrm{col})}$ for local and collective dissipation, is given by
	\begin{equation}
		f_{\mathrm{active}}^{\mathrm{loc}(\mathrm{col})} = 1-\frac{N_F}{2^N}.
	\end{equation}
	In the next section, we highlight how the metastable-like protection of the subspace $\bar{\mathcal{H}}_B$ enhances the charging process of the quantum battery by comparing both local and collective dissipation, as well as the effect of the reservoir temperature by considering the three reservoir scenarios: cold, normal, and hot baths, respectively.
	
	%*********************************************************%
	\begin{figure*}[t!]
		\centering
		\includegraphics[width=0.98\textwidth]{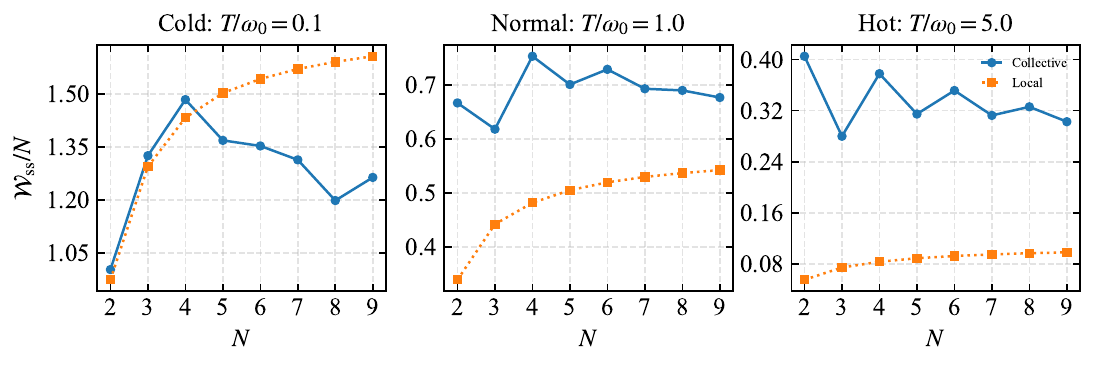}\\
		\includegraphics[width=0.98\textwidth]{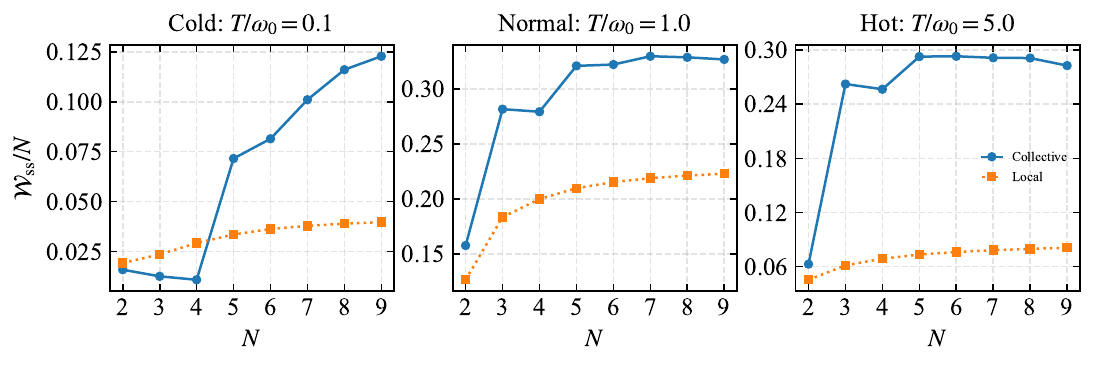}
		\caption{Stationary-state ergotropy per qubit, $\mathcal{W}_{ss}/N$, as a function of the number of qubits $N$ for three bath-temperature scenarios: cold bath ($T/\omega_0 = 0.1$), normal bath ($T/\omega_0 = 1$), and hot bath ($T/\omega_0 = 5$), respectively, for collective and local dissipation. The top panel corresponds to AFM coupling ($J=-1$), while the bottom panel corresponds to FM coupling ($J=1$). The solid red curves correspond to collective interactions, while the dotted blue curves correspond to local interactions with the bath. The remaining parameters are set to $h=0.1$, $\gamma=0.09$, and $\omega_0=2h$.}
		\label{WSS}
	\end{figure*}
	%********************************************************%
	%********************************************************************%
	\section{Impact of dark and frozen states on the charging process of the quantum battery} \label{Sec:Scaling}
	%********************************************************************%
	
	The maximal extractable work from the quantum battery is quantified by the ergotropy, defined in Eq.~\eqref{Eq.ERGOTROPY}. It is determined by the difference between the stored energy of the battery [Eq.~\eqref{Eq.Stored_Energy}] and the energy of its passive state [Eq.~\eqref{Eq.Stored_Energy_passive}]. The passive state is constructed by assigning the largest populations to the lowest-energy eigenstates, thereby minimizing the battery energy under unitary transformations. Consequently, the population distribution plays a crucial role in determining the ergotropy. In this context, the presence of dark and frozen states under collective dissipation, in contrast to their absence under local dissipation, becomes particularly important. Since these states are immune to reservoir-induced decoherence, they can significantly modify the population distribution and, consequently, the ergotropy. In the following, we investigate how dark and frozen states influence the work-extraction capability of the quantum battery under collective dissipation compared with local dissipation.
	
	%********************************************************************%
	\subsection{Ergotropy and charging power }\label{Ergotropy_power}
	%********************************************************************%
	In this section, we analyze the dynamics of the ergotropy, charging power, and stationary-state ergotropy as functions of the number of qubits to investigate the charging performance of the quantum battery. We compare collective and local dissipation for both the FM and AFM cases to reveal the role of the active Hilbert space in the charging process. We also examine the effect of the reservoir temperature by considering cold, intermediate-temperature, and hot baths. In our numerical calculations, the quantum battery is initially prepared in the ground state of the diagonalized battery Hamiltonian given in Eq.~\eqref{Diagonalized_Hamiltonian}. For both the FM and AFM cases, this ground state is passive, as illustrated in Fig.~\ref{ERGOTROPY}.
	
	In Fig.~\ref{ERGOTROPY}, we show the dynamics of the ergotropy of the quantum battery, $\mathcal{W}(t)$, as a function of time and the number of qubits $N$ for three reservoir temperatures and for the AFM (top panel) and FM (bottom panel) configurations. We observe that collective dissipation generally yields higher ergotropy than local dissipation for both FM and AFM configurations, although the relative advantage depends on the number of qubits and the reservoir temperature. Physically, in the local-dissipation regime, the bath treats each spin independently, so no cooperative processes occur. In contrast, in the collective dissipation regime, the bath couples to the total spin via the interaction operators in Eq.~\eqref{OPINT_COL}. This leads to important phenomena associated with symmetry protection through the formation of dark states, as well as metastable-like protection arising from the presence of frozen states. These effects provide an important cooperative contribution to the energy transfer and injection processes in the quantum battery. Importantly, the battery is charged autonomously in both the FM and AFM regimes, under both collective and local dissipation.
	
	To highlight the charging performance over time, we compute the maximal charging power $\mathcal{P}_{\max}$, which is defined as the maximal value of the power $\mathcal{P}(t)$ at the corresponding time $t_{\max}$. In Fig.~\ref{PMAX}, we represent $\mathcal{P}_{\max}$ as a function of the number of qubits $N$ for both FM and AFM cases, and for collective and local dissipation, under the three reservoir-temperature scenarios. As observed for ergotropy, the charging power is higher for collective dissipation than for local dissipation in both FM and AFM cases. Moreover, for collective dissipation in both FM and AFM cases, the maximal charging-power curves scale approximately linearly with the number of qubits $N$, whereas in the local case they increase much more slowly. This means that the maximal charging power in the collective case is higher than in the local case.
	
	%*********************************************************%
	\begin{figure*}[t!]
		\centering
		\includegraphics[width=0.98\textwidth]{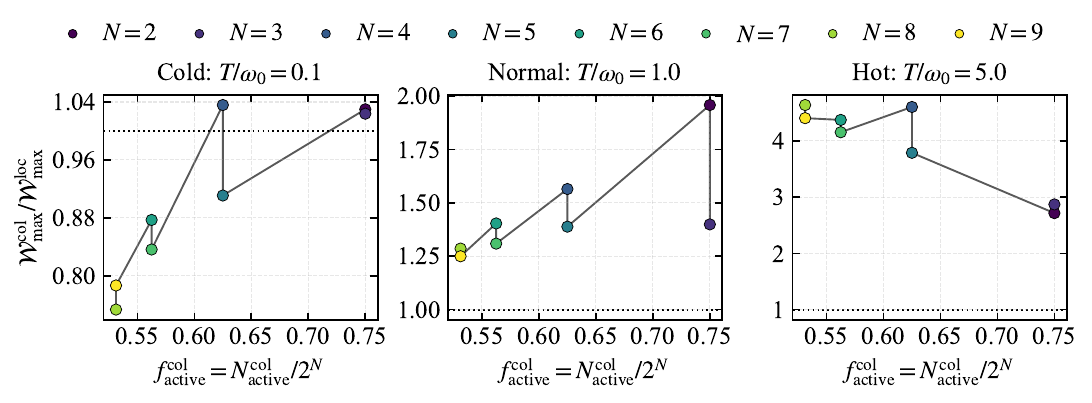}\\
		\includegraphics[width=0.98\textwidth]{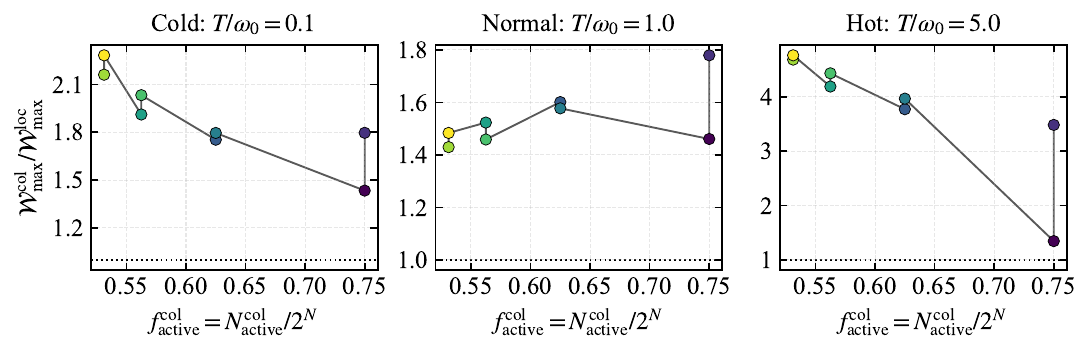}
		\caption{Enhancement factor $\mathcal{W}_{\max}^{\mathrm{col}}/\mathcal{W}_{\max}^{\mathrm{loc}}$ as a function of $f_{\mathrm{active}}^{\mathrm{col}}$ for three bath-temperature scenarios: cold bath ($T/\omega_0 = 0.1$), normal bath ($T/\omega_0 = 1$), and hot bath ($T/\omega_0 = 5$), respectively. Panel (a) corresponds to AFM coupling ($J=-1$), while panel (b) corresponds to FM coupling ($J=1$). The remaining parameters are set to $B=0.1$, $\gamma=0.09$, and $\omega_0=2B$.}
		\label{ENHANCMENT_FACTROR}
	\end{figure*}
	%********************************************************%
	After the quantum battery reaches its stationary state under collective and local dissipation, we investigate how the stationary-state ergotropy depends on the number of qubits $N$. This quantity, denoted by $\mathcal{W}_{ss}/N$, represents the contribution of each qubit to the ergotropy at the stationary state, as shown in Fig.~\ref{WSS}. At the stationary state, the ergotropy contribution of each qubit increases with the number of qubits in the FM case and decreases in the AFM case. A similar behavior is observed for local dissipation, while the relative advantage of collective dissipation depends on the bath temperature, magnetic configuration, and system size. In most parameter regimes considered here, collective dissipation enhances the stationary-state ergotropy, although exceptions can occur, particularly in the low-temperature AFM regime for larger $N$.
	
	However, the results show that collective dissipation can provide a substantial advantage over local dissipation, depending on the bath temperature, magnetic configuration, and system size. In particular, the advantage is pronounced in the normal- and hot-temperature regimes, whereas in the cold AFM regime the relative performance of the two dissipation mechanisms can become comparable or may even favor local dissipation for sufficiently large $N$. Note that the reservoir temperature plays an important role in the charging performance of the quantum battery for both collective and local dissipation. For the cold bath, the probability of absorption from the reservoirs is much smaller than that of emission, $\gamma_{\uparrow}\ll \gamma_{\downarrow}$, and the total ergotropy increases with $N$. However, for the AFM case, at low temperature, almost all the population initially occupies the ground state. Only a limited number of excitations can be extracted from the bath, and these excitations are distributed among more qubits; therefore, $\mathcal{W}_{ss}/N$ decreases. This behavior instead reflects the interplay between the system size, magnetic configuration, and the fixed engineered dissipative rates. For the FM case, the decrease is slighter, and eventually the per-qubit ergotropy stabilizes. This suggests that ferromagnetic ordering allows collective excitations to spread more efficiently over the battery.
	
	For the normal bath, the probability of energy absorption from the bath is smaller than that of emission, $\gamma_{\uparrow}\leq \gamma_{\downarrow}$, and this regime generally provides a favorable balance between energy injection and dissipative noise. Consequently, collective dissipation can efficiently enhance the stationary-state ergotropy, although the magnitude of this enhancement depends on the sign of $J$ and $N$. The bath contains enough thermal excitations, while the thermal noise is still moderate. Consequently, collective coherence builds up efficiently, allowing cooperative charging without excessive thermal disorder. This temperature represents the best compromise between energy availability and quantum coherence.
	For the hot bath, the probability of energy absorption from the bath is similar to that of emission, $\gamma_{\uparrow}\sim\gamma_{\downarrow}$. Thermal fluctuations simultaneously destroy coherence, so the system relaxes toward a mixed steady state. Therefore, after the initial transient, the ergotropy becomes nearly constant, and the advantage of collective dissipation becomes relatively weak in this high-temperature regime.
	
	A remarkable difference between the two magnetic configurations is observed in the achievable ergotropy. In particular, the AFM configuration ($J<0$) exhibits a substantially larger ergotropy than the FM configuration ($J>0$) throughout the charging dynamics, especially in the low-temperature regime, where the maximum ergotropy reaches values close to $16$, whereas the FM case remains below approximately $3$; see Fig.~\ref{ERGOTROPY}. This pronounced enhancement originates from the different organization of the protected Hilbert-space sectors within the many-body energy spectrum. It is important to emphasize that, for a given system size, both the AFM and FM configurations possess the same numbers of dark states and frozen states. The crucial difference lies in their energetic distribution rather than in their multiplicity. As shown in Appendix~\ref{appendix.dark_stat and frozen state position}, the protected states occupy different regions of the battery energy spectrum in the two magnetic phases. In the AFM configuration, the dark and frozen states are predominantly located in the low-energy sector, where they effectively suppress dissipative losses, preserve quantum coherence, and protect the stored energy over long time scales. Consequently, a larger fraction of the absorbed energy remains extractable as ergotropy. In contrast, although the FM interaction favors collective spin alignment and enhances superradiant energy exchange, the protected states are shifted to different energy sectors, reducing their ability to preserve the useful stored energy against dissipation. These results demonstrate that the charging advantage of the AFM configuration is not determined by the number of dark or frozen states, but rather by their position within the battery energy spectrum, which governs how effectively the metastable-like subspaces contribute to the charging process.
	
	%********************************************************************%
	\subsection{Effect of metastable-level protection on the charging process}
	%********************************************************************%
	
	As discussed above, collective dissipation is better than local dissipation for enhancing the ergotropy of the quantum battery. The main difference between the two types of dissipation is the presence of a metastable-like protected subspace in the case of collective dissipation. We also observe that temperature is an important resource for the performance of the charging process. Here, we analyze the effect of the active Hilbert space on the charging process and how metastable protection enhances the charging process for both FM and AFM cases, considering three reservoir-temperature scenarios. We focus on the enhancement factor, denoted by $\mathcal{W}_{\max}^{\mathrm{col}}/\mathcal{W}_{\max}^{\mathrm{loc}}$, where $\mathcal{W}_{\max}^{\mathrm{col}(\mathrm{loc})}$ quantifies the maximal ergotropy over time, namely
	\begin{equation}
		\mathcal{W}_{\max}^{\mathrm{col}(\mathrm{loc})}
		=
		\max_t \mathcal{W}(t),
	\end{equation}
	for the collective or local case. We analyze this enhancement factor $\mathcal{W}_{\max}^{\mathrm{col}}/\mathcal{W}_{\max}^{\mathrm{loc}}$ as a function of $f_{\mathrm{active}}^{\mathrm{col}}$ in Fig.~\ref{ENHANCMENT_FACTROR}, to highlight the role of the active levels in the Hilbert space in maximal ergotropy production. This is done for both FM and AFM cases and for different reservoir temperatures. As discussed in Sec.~\ref{Ergotropy_power}, for the AFM case, the frozen sector still protects the stored energy, but the reduction of active charging pathways partially compensates this advantage. Therefore, collective dissipation generally remains beneficial, although its advantage progressively weakens as the active charging pathways are reduced. This means that protection and charging efficiency compete with each other. For the FM case, the collective metastable sector strongly protects the battery against thermal dissipation, whereas the local bath rapidly destroys the extractable work. As a result, collective dissipation can provide a substantial advantage, particularly at high temperature, although its magnitude depends on the system parameters. Moreover, the largest collective advantage occurs at high temperature.
	
	Physically, the size of the metastable-like protected subspace is not sufficient to determine the charging performance. Although the FM and AFM cases have the same contribution of the active Hilbert space to the interaction, $f_{\mathrm{active}}^{\mathrm{col}}$, they exhibit different ergotropy enhancements due to the arrangement of the frozen states in the many-body energy spectrum, which affects the pathways of energy injection into the quantum battery. In other words, metastable protection is not necessary for enhancing ergotropy production, but it is sufficient.

	%********************************************************%
	\section{Conclusion}\label{conc}
	%********************************************************%
	We investigated the influence of decoherence-free and metastable-protected subspaces on the charging performance of open quantum batteries described by the transverse-field Ising model. By comparing local and collective dissipation, we demonstrated that collective coupling to a common reservoir gives rise to symmetry-protected dark states together with a significantly larger set of frozen states, which form an extended metastable-like protected subspace. We derived the exact number of dark states analytically from the multiplicity of the singlet sector, showing that it follows the Catalan sequence for even system sizes, whereas frozen states emerge for both even and odd numbers of qubits.
	
	Our analysis shows that collective dissipation can enhance both the ergotropy and charging power relative to local dissipation, with the advantage depending on the system parameters. More importantly, we showed that the charging advantage cannot be explained solely by the number of protected states. Instead, the decisive factor is their location within the many-body energy spectrum. Although the FM and AFM configurations possess identical numbers of dark and frozen states, the AFM phase exhibits substantially larger extractable work because the protected states are preferentially located in low-energy regions where they more efficiently suppress dissipative losses and preserve useful stored energy.
	
	Furthermore, by introducing the active Hilbert-space fraction, we demonstrated that the charging dynamics are governed by the competition between the protection of stored energy and the dissipative connectivity of the active Hilbert space. The metastable-like protected subspace reduces irreversible energy losses while maintaining sufficient active pathways for the charging process, thereby providing an effective mechanism for enhancing the performance of the quantum battery. These findings identify the dark-state and frozen-state sectors as valuable resources for autonomous quantum batteries and demonstrate that engineering the symmetry and spectral structure of open quantum systems represents a promising strategy for designing robust and high-performance quantum energy-storage devices.
	
	\section*{Acknowledgments}
	A.~K. and A.~U. contributed equally to this work.
	A.~K. acknowledges the CNRST–Morocco for financial support through the program ``PhD-Associate Scholarship – PASS'' and the hospitality of Prof.~Ö.~E.~M. and his QuEST research group at the Department of Physics, Koç University, where part of this work was carried out.

	\appendix
	\section{Derivation of the Dark States and Catalan Multiplicity}
	\subsection{Definition and Hilbert Space Decomposition}
	
	Dark states are defined as the quantum states $|\psi_0\rangle$ that are completely annihilated by the collective ladder operators $J_{\pm}$:
	\begin{equation}
		J{\pm} |\psi_0\rangle = 0, \quad J^\dagger |\psi_0\rangle = 0,
	\end{equation}
	where the collective spin operators for a system of $N$ qubits are defined as:
	\begin{equation}
		J_{\pm} = \sum_{i=1}^{N} \sigma_{i}^{\pm}, \quad J_{\beta} = \sum_{i=1}^{N} \sigma_{i}^{\beta} \quad \text{for } \beta \in \{x,y,z\}.
	\end{equation}
	The total squared spin operator is given by $J^2 = \sum_{\beta} J_{\beta}^2$. 
	
	The complete Hilbert space $\mathcal{H}_B$ for $N$ qubits can be decomposed into a direct sum over total spin sectors $S$ via the decomposition:
	\begin{equation}\label{DEC_Hilbert}
		\mathcal{H}_B = \bigoplus_{S} \mathcal{H}_S \otimes \mathcal{M}_S = \bigoplus_{S} d_S \mathcal{H}_S,
	\end{equation}
	where $\mathcal{H}_S$ represents the spin representation space, $\mathcal{M}_S$ is the degeneracy or multiplicity space, and $d_S = \dim[\mathcal{M}_S]$ corresponds to the multiplicity of the total spin $S$.
	
	Any arbitrary state $|\Psi\rangle$ can be expanded in the total spin basis $|S, M, \alpha\rangle$ as:
	\begin{equation}\label{anypsi}
		|\Psi\rangle = \sum_{S, M, \alpha} a_{S,M,\alpha} |S, M, \alpha\rangle,
	\end{equation}
	subject to the normalization condition $\sum_{S,M,\alpha} |a_{S,M,\alpha}|^2 = 1$. Here, $S$ is the total spin quantum number, $M$ is the magnetic quantum number (ranging from $-S \leq M \leq S$), and $\alpha = 1, 2, \dots, d_S$ is the multiplicity index. 
	
	Using the Clebsch-Gordan coefficients, we can transition from the total spin basis to the standard computational basis via:
	\begin{equation}
		|S,M,\alpha\rangle = \sum_{m_1, \dots, m_N} C_{m_1, \dots, m_N}^{S,M,\alpha} |m_1, \dots, m_N\rangle,
	\end{equation}
	where $M = \sum_{i=1}^{N} m_i$ and $m_i = \pm \frac{1}{2}$ represent the local magnetic quantum numbers.
	\subsection{Proof of the Dark State Condition}
	
	The actions of the total spin operators on our basis states $|S,M,\alpha\rangle$ (setting $\hbar = 1$) yield:
	\begin{align}
		J^2 |S,M,\alpha\rangle &= S(S+1)|S,M,\alpha\rangle, \\
		J_z |S,M,\alpha\rangle &= M|S,M,\alpha\rangle, \\
		J_{\pm} |S,M,\alpha\rangle &= \sqrt{S(S+1) - M(M \pm 1)} |S, M \pm 1, \alpha\rangle.
	\end{align}
	
	To satisfy the dark state criterion $J_{\pm}|\psi\rangle = 0$, we substitute our state expansion into the ladder operator eigenvalue relation:
	\begin{equation}
		J_{\pm}|\psi\rangle = \sum_{S,M,\alpha} a_{S,M,\alpha} \sqrt{S(S+1) - M(M \pm 1)} |S, M \pm 1, \alpha\rangle = 0.
	\end{equation}
	
	For this to vanish identically, the coefficients must satisfy:
	\begin{align}
		S(S+1) - M(M+1) = 0 &\implies M = S, \\
		S(S+1) - M(M-1) = 0 &\implies M = -S.
	\end{align}
	Combining both conditions, simultaneous annihilation by $J_+$ and $J_-$ requires
	\begin{equation}\label{SM}
		S = M = 0.
	\end{equation}
	Thus, the dark states are precisely the states belonging to the total-spin singlet sector.
	For a system of $N$ qubits, the allowed values for total spin decrease in integer steps from the maximum value:
	\begin{equation}
		S \in \left\{ \frac{N}{2}, \frac{N}{2}-1, \frac{N}{2}-2, \dots, S_{\min} \right\}.
	\end{equation}
	Therefore, achieving a dark state requires $S_{\min} = 0$, which is only achievable if the number of qubits is even, i.e., $N = 2n$ where $n \in \mathbb{N}^*$. If $N$ is odd, $S_{\min} = \frac{1}{2}$, so no $S = 0$ sector exists and hence no dark states satisfying $J_+ |\psi\rangle = J_- |\psi\rangle = 0$ can occur.		
	\subsection{Derivation of Dark State Multiplicity via Catalan Numbers}
	\label{Der_Dark_state}
	The total dimension of a given magnetization sector $M$ in the computational basis is denoted by $g_M$. Let $N_{\uparrow}$ be the number of spins pointing up ($m_i = \frac{1}{2}$) and $N_{\downarrow}$ be the number of spins pointing down ($m_i = -\frac{1}{2}$), where $N = N_{\uparrow} + N_{\downarrow}$. The total magnetization is:
	\begin{equation}
		M = \frac{1}{2}(N_{\uparrow} - N_{\downarrow}) = \frac{N}{2} - N_{\downarrow} \implies N_{\downarrow} = \frac{N}{2} - M.
	\end{equation}
	
	The number of computational basis configurations possessing a specific total magnetization value $M$ is:
	\begin{equation}
		g_M = \binom{N}{\frac{N}{2} - M}.
	\end{equation}
	
	From the decomposition of the total Hilbert space, the dimension can also be written as a sum over the multiplicities of the contributing total spin sectors:
	\begin{equation}
		g_M = \sum_{S \geq |M|} d_S.
	\end{equation}
	By exploiting the symmetry property $g_{M} = g_{-M}$, we can isolate individual spin multiplicities by taking the difference between consecutive magnetization sectors:
	\begin{align}
		g_M &= \sum_{S=M}^{N/2} d_S, \\
		g_{M+1} &= \sum_{S=M+1}^{N/2} d_S.
	\end{align}
	Subtracting these two equations yields the general formula for the multiplicity $d_S$:
	\begin{equation}\label{ds_multi}
		d_S = g_S - g_{S+1} = \binom{N}{\frac{N}{2} - S} - \binom{N}{\frac{N}{2} - S - 1}.
	\end{equation}
	
	The total number of dark states $N_D$ corresponds directly to the multiplicity of the $S = 0$ sector space ($d_0$). Setting $S = 0$ and substituting $N = 2n$, we get:
	\begin{equation}\label{DSNAP}
		N_D = d_0 = \binom{2n}{n} - \binom{2n}{n-1}.
	\end{equation}
	
	Expanding the binomial coefficients algebraically reveals the final relationship:
	\begin{align}\label{CATALAN}
		N_D &= \frac{(2n)!}{n!n!} - \frac{(2n)!}{(n-1)!(n+1)!} \nonumber \\
		&= \frac{(2n)!}{n!(n+1)!} \left[ (n+1) - n \right] \nonumber \\
		&= \frac{1}{n+1}\binom{2n}{n} = C_n,
	\end{align}
	where $C_n$ is the $n$-th \textbf{Catalan number}. This proves that the number of dark states available in a collective system of $2n$ qubits scales exactly with the Catalan sequence.
	%*********************************************************************%
	\section{Position of the frozen and dark States in the FM and AFM configurations: The Case of $N=2$} \label{appendix.dark_stat and frozen state position}
	%*********************************************************************%
	For the two-qubit case ($N=2$), the battery Hamiltonian $H_B$ can be diagonalized exactly. We denote its eigenstates and eigenvalues by $\{|E_k^{\alpha}\rangle\}$ and $\{E_k^{\alpha}\}$, respectively, where $\alpha \in \{\mathrm{AFM},\mathrm{FM}\}$ refers to the AFM regime ($J<0$) or the FM regime ($J>0$). The Hamiltonian can then be written as
	\begin{equation}
		H_B^{\alpha}
		=
		\sum_{k=0}^{3}
		E_k^{\alpha}
		|E_k^{\alpha}\rangle
		\langle E_k^{\alpha}|.
	\end{equation}
	
	The corresponding eigenvalues are given by
	\begin{equation}
		E_0^{\alpha}=-\sqrt{4h^2+J^2},
		\quad
		E_1^{\alpha}=-J,
		\quad
		E_2^{\alpha}=J,
		\quad
		E_3^{\alpha}=\sqrt{4h^2+J^2}.
	\end{equation}
	
	We note that the eigenvalues are identical in the AFM and FM cases, i.e.,
	$E_k^{\mathrm{AFM}}=E_k^{\mathrm{FM}}$. The corresponding eigenstates
	$\{|E_k^\alpha\rangle\}$ are obtained by diagonalizing $H_B$ in the
	computational basis
	$\{|\uparrow\uparrow\rangle,|\uparrow\downarrow\rangle,
	|\downarrow\uparrow\rangle,|\downarrow\downarrow\rangle\}$.
	where the normalization factors are
	\begin{align}
		\mathcal{N}_{-}
		&=
		\sqrt{
			2+
			\frac{1}{2}
			\left|
			\frac{J-\sqrt{4h^2+J^2}}{h}
			\right|^2
		},
		\\
		\mathcal{N}_{+}
		&=
		\sqrt{
			2+
			\frac{1}{2}
			\left|
			\frac{J+\sqrt{4h^2+J^2}}{h}
			\right|^2
		}.
	\end{align}
	
	Equivalently, in the computational basis
	\begin{equation}
		\left\{
		\ket{\uparrow\uparrow},
		\ket{\uparrow\downarrow},
		\ket{\downarrow\uparrow},
		\ket{\downarrow\downarrow}
		\right\},
	\end{equation}
	the normalized eigenstates can be written as
	\begin{align}
		\ket{E_1}
		&=
		\frac{1}{\sqrt{2}}
		\left(
		-\ket{\uparrow\uparrow}
		+
		\ket{\downarrow\downarrow}
		\right),
		\\
		\ket{E_2}
		&=
		\frac{1}{\sqrt{2}}
		\left(
		-\ket{\uparrow\downarrow}
		+
		\ket{\downarrow\uparrow}
		\right),
		\\
		\ket{E_3}
		&=
		\frac{1}{\mathcal{N}_{-}}
		\left[
		\ket{\uparrow\uparrow}
		-
		\frac{J-\sqrt{4h^2+J^2}}{2h}
		\left(
		\ket{\uparrow\downarrow}
		+
		\ket{\downarrow\uparrow}
		\right)
		+
		\ket{\downarrow\downarrow}
		\right],
		\\
		\ket{E_4}
		&=
		\frac{1}{\mathcal{N}_{+}}
		\left[
		\ket{\uparrow\uparrow}
		-
		\frac{J+\sqrt{4h^2+J^2}}{2h}
		\left(
		\ket{\uparrow\downarrow}
		+
		\ket{\downarrow\uparrow}
		\right)
		+
		\ket{\downarrow\downarrow}
		\right].
	\end{align}
	
	%*****************************************************************************%
	\begin{figure*}
		\centering
		\subfloat[\label{AFM_ENERGY_Sch}]{%
			\includegraphics[width=0.4\textwidth]{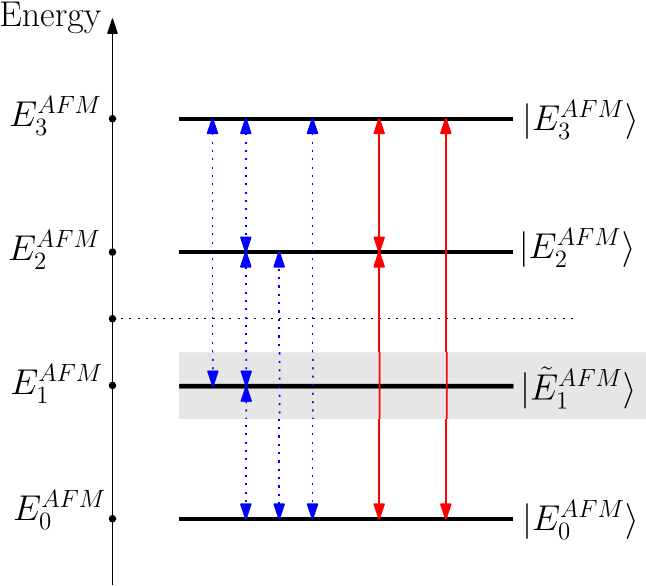}
		}\hfill
		\subfloat[\label{FM_ENERGY_Sch}]{%
			\includegraphics[width=0.4\textwidth]{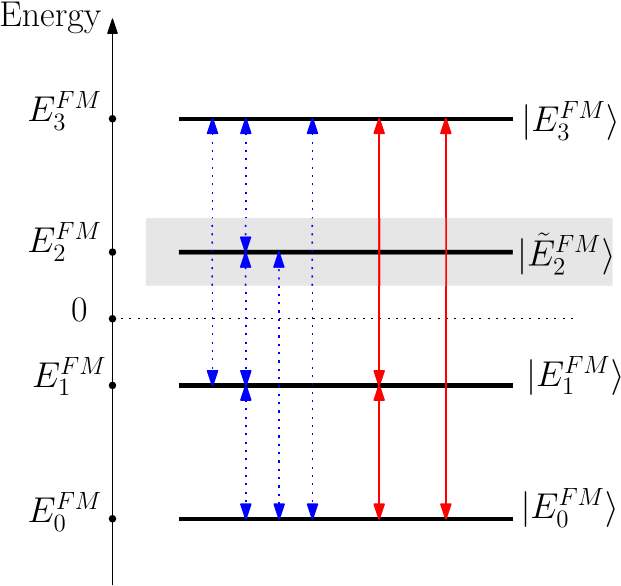}
		}
		\caption{Energy level diagrams and transition pathways for the (a) AFM and (b) FM configurations. Blue dotted lines represent the transition channels under local dissipation while red solid lines denote the transitions under collective dissipation. The shaded regions highlight the energy levels $|\tilde{E}_1^{\mathrm{AFM}}\rangle$ and $|\tilde{E}_2^{\mathrm{FM}}\rangle$, which identify the decoherence-free subspaces emerging under collective coupling. The absence of specific red transitions to these manifolds illustrates the shielding effect that suppresses energy decay and enables robust energy storage.}
		\label{fig:placeholder}
	\end{figure*}
	%*****************************************************************************%
	
	%***********************************************%
	%--------------------------------------------------%
	\begin{figure}[t!]
		\subfloat[\label{ENERGY_AFM}]{%
			\includegraphics[width=0.45\textwidth]{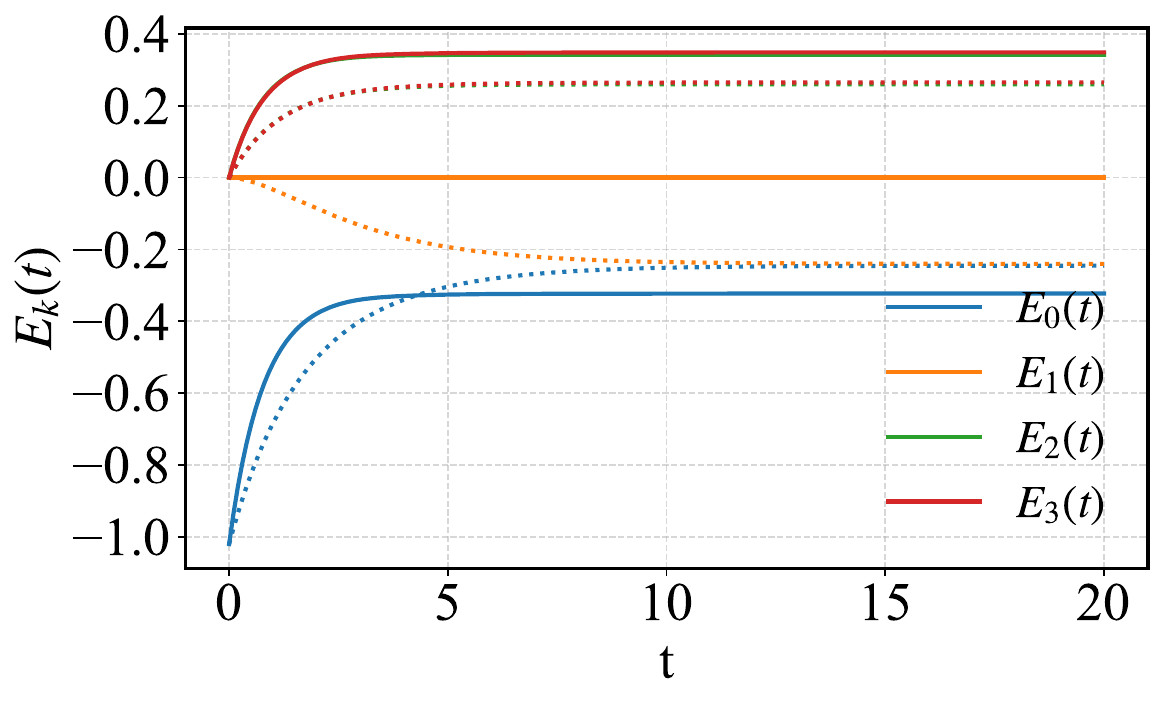}
		}
		\hfill
		\subfloat[\label{ENERGY_FM}]{%
			\includegraphics[width=0.45\textwidth]{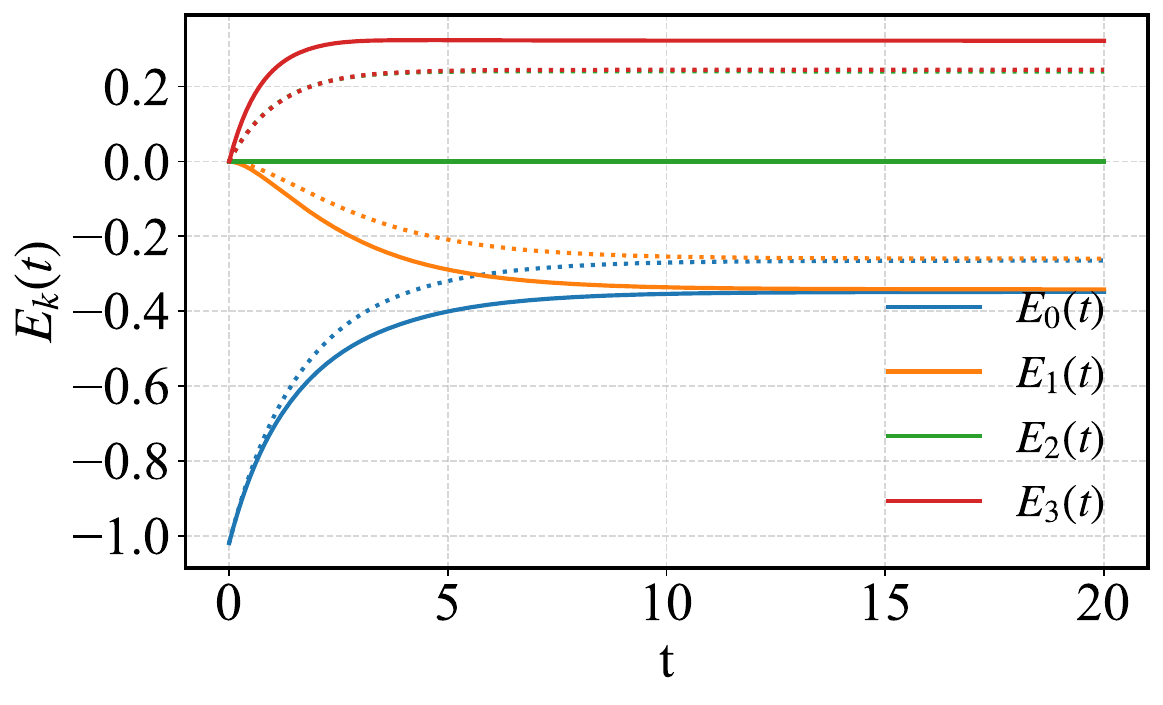}
		}
		\caption{
			Dynamics of $E_k^{\alpha}(t)$ as a function of time $t$ for
			$k=\{0,1,2,3\}$ in the (a) AFM and (b) FM regimes.
			Solid lines represent collective dissipation, whereas dotted
			lines represent local dissipation for each energy level $k$.
			The remaining parameters are set to $h=0.1$, $\gamma=0.09$,
			$\omega_0=2h$, and $T=0.5$.
		}
		\label{ENERGY}
	\end{figure}
	%--------------------------------------------------%
	%****************************************************************************%
	
	In this section, we numerically analyze the effects of local and collective
	dissipation on the charging process of the quantum battery. We also highlight
	the role of dark states under collective dissipation and the absence of
	DFSs under local dissipation in both the AFM and
	FM regimes.
	
	\subsection{Energy dynamics under local and collective dissipation: Role of dark states}
	
	The energy stored in the quantum battery can be expressed as a sum over all
	energy levels $k$:
	\begin{align}
		E(\rho(t))
		&=
		\sum_{k=0}^{2^N-1} E_k^\alpha(t),
		\\
		E_k^{\alpha}(t)
		&=
		P_k^\alpha(t)E_k^{\alpha}.
	\end{align}
	Because the Hamiltonian is time independent, the dynamics of $E(\rho(t))$ is
	fully determined by the evolution of the state $\rho(t)$ through the
	populations $P_k^\alpha(t)$.
	
	Collective dissipation in our model is described by the dissipator
	$\mathcal{L}_{\mathrm{col}}[\rho]$, which arises from a
	permutation-symmetric system--bath coupling mediated by the collective jump
	operator
	\begin{equation}
		L=\sum_{i=1}^{N}L_i.
	\end{equation}
	This symmetry reflects the indistinguishability of the decay channels
	associated with the different qubits and constrains the dissipative structure
	of the dynamics. Consequently, the evolution generated by
	$\mathcal{L}_{\mathrm{col}}[\rho]$ preserves a decoherence-free subspace that
	remains invariant under the dissipative dynamics. States belonging to this
	subspace,
	$\ket{\tilde{E}_k^{\alpha}}\in\tilde{\mathcal{H}}$, lie in the kernel of the
	collective jump operator and are therefore annihilated by the dissipator,
	thereby constituting dark states of the evolution. Dissipative transitions
	are consequently restricted to states orthogonal to the DFS.
	
	In contrast, under local dissipation, each qubit interacts independently with
	its own environment through the local jump operators $L_i$. This independent
	coupling breaks the permutation symmetry required for the formation of a DFS
	and therefore prevents the emergence of symmetry-protected subspaces. As a
	result, no decoherence-free subspace is present, and the system eigenstates
	are generally affected by dissipative decay.
	
	The initial state considered in our analysis is given by
	\begin{equation}
		\label{ini_state}
		\rho^{\alpha}(0)
		=
		\ket{E_0^\alpha}\bra{E_0^\alpha}.
	\end{equation}
	
	The energy-level structure and transition pathways for $N=2$ are depicted in
	Fig.~\ref{fig:placeholder}. Under collective dissipation, represented by the
	red solid lines, the system exhibits a restricted transition topology. In the
	AFM regime, shown in panel (a), the environment induces transitions
	exclusively within the
	$\{\ket{E_0},\ket{E_2},\ket{E_3}\}$ manifold, leaving
	$\ket{\tilde{E}_1^{\mathrm{AFM}}}$ as a decoherence-free dark state.
	Conversely, in the FM regime, shown in panel (b), the state
	$\ket{\tilde{E}_2^{\mathrm{FM}}}$ is decoupled from the collective bath.
	This change in the position of the dark state reflects the eigenvalue
	reordering governed by the sign of the exchange coupling $J$.
	
	In contrast, local dissipation, represented by the blue dotted lines, allows
	transitions between all energy levels because the independent coupling of
	each qubit to its respective reservoir breaks the symmetry required for state
	protection. This comparison demonstrates the essential role of collective
	effects in protecting stored energy against environmental dissipation.
	
	%*********************************************************%
	\begin{figure}[t!]
		\subfloat[\label{ERGO_AFM_app}]{%
			\includegraphics[width=0.45\textwidth]{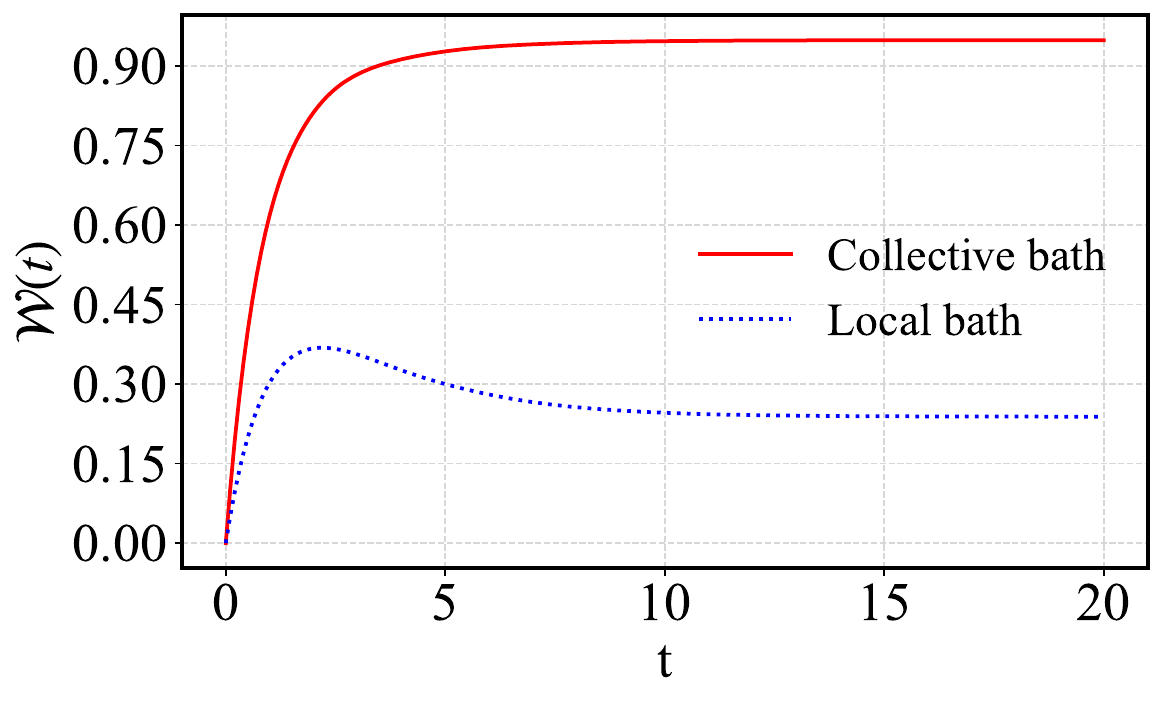}
		}
		\hfill
		\subfloat[\label{ERGO_FM_app}]{%
			\includegraphics[width=0.45\textwidth]{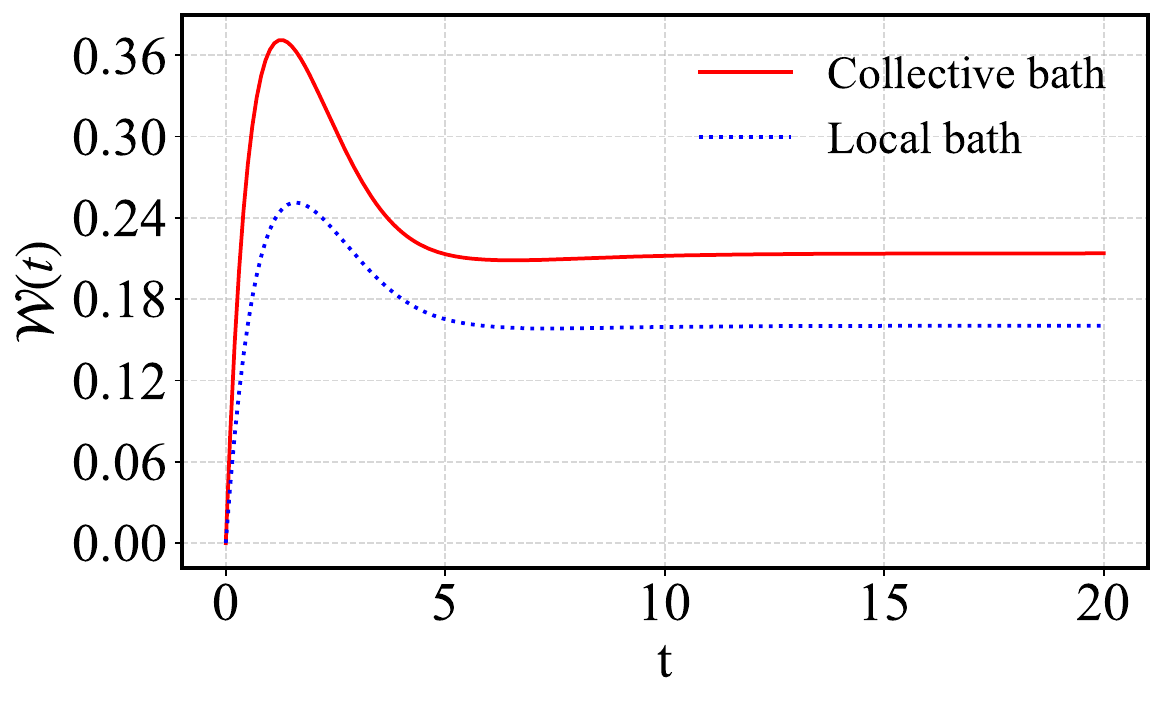}
		}
		\caption{
			Dynamics of the ergotropy $\mathcal{W}(\rho(t))$ as a function of
			time $t$ for the (a) AFM ($J=-1$) and (b) FM ($J=1$) regimes.
			The solid red curve corresponds to collective dissipation, whereas
			the dotted blue curve corresponds to local dissipation.
			The remaining parameters are set to $h=0.1$, $\gamma=0.09$,
			$\omega_0=2h$, and $T=0.5$.
		}
		\label{ERGOTROPY_app}
	\end{figure}
	%********************************************************%
	
	Figure~\ref{ENERGY} illustrates the dissipative time evolution of the
	level-resolved energies $E_k^\alpha(t)$ for the initial state
	$\rho^{\alpha}(0)=\ket{E_0^\alpha}\bra{E_0^\alpha}$. Solid lines correspond
	to collective dissipation, whereas dotted lines represent local dissipation.
	In both the AFM and FM regimes, local dissipation results in relaxation across
	all energy levels, reflecting the absence of symmetry-induced selection
	rules in the transition dynamics. The increase in $E_0^\alpha(t)$ reflects
	the depletion of the corresponding population $P_0^\alpha(t)$ due to thermal
	excitation, accompanied by a redistribution of the population toward
	higher-energy levels.
	
	Under collective dissipation in the AFM regime
	[Fig.~\ref{ENERGY_AFM}], $E_0^{\mathrm{AFM}}(t)$ increases as
	$P_0^{\mathrm{AFM}}(t)$ decreases, while
	$E_2^{\mathrm{AFM}}(t)\approx E_3^{\mathrm{AFM}}(t)$. This behavior reflects
	the symmetric transition pathways imposed by the permutation-symmetric
	system–bath coupling. The dark-state contribution
	$E_1^{\mathrm{AFM}}(t)$ remains dynamically frozen because of its decoupling
	from the dissipative channels.

	\subsection{Impact of dark states on work extraction: Comparison between local and collective dissipation}
	
	The extractable work from a quantum battery is quantified by the difference between its mean energy, $E(\rho(t))$, and the energy of the corresponding passive state, $E_{\mathrm{passive}}(\rho(t))$. Under collective dissipation, the emergence of dark states modifies the system dynamics by suppressing specific transition channels between energy levels. This behavior contrasts with local dissipation, for which such constraints are absent. Since the passive state is determined by the population distribution in the energy eigenbasis, we analyze the impact of dark states on work extraction in both the AFM and FM configurations by comparing collective and local dissipation.
	
	Figure~\ref{ERGOTROPY_app} shows the time evolution of the ergotropy for the initial state given in Eq.~\eqref{ini_state}, in both the AFM [Fig.~\ref{ERGO_AFM_app}] and FM [Fig.~\ref{ERGO_FM_app}] regimes. In both cases, the ergotropy $\mathcal{W}(\rho(t))$ is consistently higher under collective dissipation than under local dissipation. It increases from its initial value over time, indicating a progressive increase in the usable work generated by the reservoir-induced dissipative dynamics.
	
	%*******************************************************%
	\begin{figure}[t!]
		\centering
		\includegraphics[width=\linewidth]{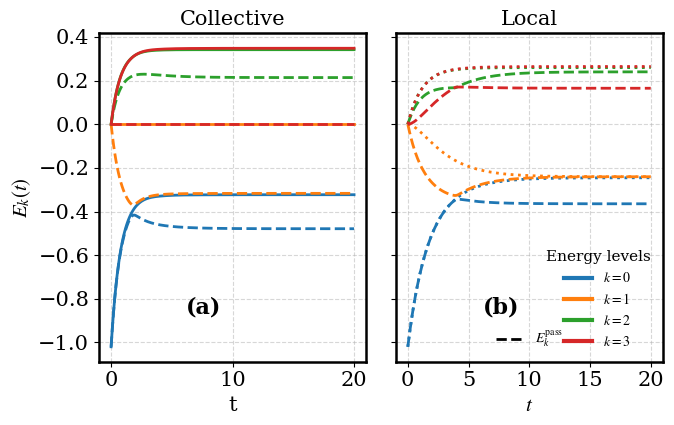}
		\includegraphics[width=\linewidth]{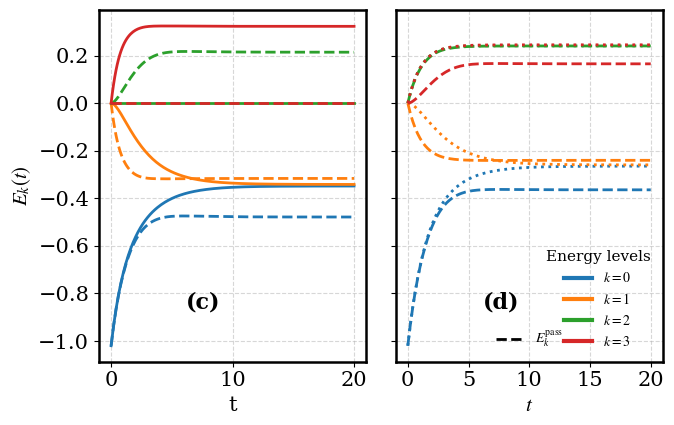}
		\caption{
			Comparison between the energy dynamics of the state and its corresponding passive state. The left column [(a), (c)] corresponds to collective dissipation, whereas the right column [(b), (d)] corresponds to local dissipation. The top row [(a), (b)] represents the AFM regime, while the bottom row [(c), (d)] represents the FM regime. Solid and dashed lines denote the energies of the state and the passive state, respectively, for the different energy levels $k$. The remaining parameters are set to $h=0.1$, $\gamma=0.09$, $\omega_0=2h$, and $T=0.5$.
		}
		\label{COMP}
	\end{figure}
	%********************************************************%
	
	Collective dissipation enhances the extractable work of the quantum battery by modifying the structure of the dissipative transitions. In the AFM regime, both the population and coherence contributions play significant roles in the charging process. In contrast, in the FM regime, the population contribution dominates at short times but vanishes in the long-time limit, leaving coherence as the primary steady-state contribution.
	
	To clarify the mechanism responsible for this behavior, Fig.~\ref{COMP} presents the dynamics of the energy-level contributions of the state and its passive counterpart, thereby highlighting the role of dark states in enhancing work extraction.
	
	Under local dissipation, all energy levels are coupled to the reservoirs in both the AFM and FM regimes. For the diagonal states considered here, the contribution of each energy level k to the ergotropy is given by
	\begin{equation}
		\label{E_D_S}
		\Delta\mathcal{W}_k^{\alpha}(t)
		=
		\left[
		P_k^{\alpha}(t)
		-
		P_{k,\mathrm{passive}}^{\alpha}(t)
		\right]
		E_k^{\alpha},
	\end{equation}
	where
	\begin{equation}
		\mathcal{W}(t)
		=
		\sum_{k=0}^{2^N-1}
		\Delta\mathcal{W}_k^{\alpha}(t).
	\end{equation}
	This level-resolved expression is used only for states diagonal in the energy basis; the general ergotropy is defined through the spectral decomposition of the density matrix. In the AFM case, the dark state $\ket{\tilde{E}_1^{\mathrm{AFM}}}$ remains completely decoupled from the dissipative dynamics. Consequently, its population remains constant in time:
	\begin{equation}
		P_1^{\mathrm{AFM}}(t)
		=
		P_1^{\mathrm{AFM}}(0).
	\end{equation}
	Similarly, in the FM case, the dark state $\ket{\tilde{E}_2^{\mathrm{FM}}}$ remains unaffected by dissipation, and its population satisfies
	\begin{equation}
		P_2^{\mathrm{FM}}(t)
		=
		P_2^{\mathrm{FM}}(0).
	\end{equation}
	
	For the passive state, the population redistribution leads to
	\begin{equation}
		P_{3,\mathrm{passive}}^{\mathrm{AFM}}(t)=0,
		\qquad
		P_{3,\mathrm{passive}}^{\mathrm{FM}}(t)=0,
	\end{equation}
	indicating that the population of level $k=3$ vanishes in the corresponding passive-state distribution.
	
	Accordingly, the relevant contributions can be written as
	\begin{align}
		\Delta\mathcal{W}_1^{\mathrm{AFM}}(t)
		&=
		P_3^{\mathrm{AFM}}(t)E_1^{\alpha},
		\\
		\Delta\mathcal{W}_2^{\mathrm{FM}}(t)
		&=
		P_3^{\mathrm{FM}}(t)E_2^{\alpha},
		\\
		\Delta\mathcal{W}_3^{\alpha}(t)
		&=
		P_3^{\alpha}(t)E_3^{\alpha}.
	\end{align}
	
	These relations indicate that dark states effectively isolate specific energy levels from the dissipative channels, thereby allowing them to retain stored energy. Finally, the vanishing of the population contribution to the ergotropy in the FM regime, under both local and collective dissipation, can be attributed to a transient population inversion relative to the passive-state ordering. For $t\lesssim 7.5$, the population distribution differs from its passive counterpart, for example,
	\begin{equation}
		E_1^{\mathrm{FM}}(t)
		\geq
		E_{1,\mathrm{passive}}^{\mathrm{FM}}(t),
	\end{equation}
	resulting in a positive contribution to the ergotropy. For $t\gtrsim 7.5$, the populations approach the passive ordering, for example,
	\begin{equation}
		E_1^{\mathrm{FM}}(t)
		\leq
		E_{1,\mathrm{passive}}^{\mathrm{FM}}(t),
	\end{equation}
	%************************************************************************************%
    \bibliography{References}
	%************************************************************************************%	
	
\end{document}